

\magnification 1200

%

%
\font\eightrm=cmr8
\font\eighti=cmmi8
\font\eightsy=cmsy8
\font\eightbf=cmbx8
\font\eighttt=cmtt8
\font\eightit=cmti8
\font\eightsl=cmsl8
\font\sixrm=cmr6
\font\sixi=cmmi6
\font\sixsy=cmsy6
\font\sixbf=cmbx6
\catcode`@11
\newskip\ttglue
\font\grrm=cmbx10 scaled 1200

\def\eightpoint{\def\rm{\fam0\eightrm}
\textfont0=\eightrm \scriptfont0=\sixrm \scriptscriptfont0=\fiverm
\textfont1=\eighti \scriptfont1=\sixi \scriptscriptfont1=\fivei
\textfont2=\eightsy \scriptfont2=\sixsy \scriptscriptfont2=\fivesy
\textfont3=\tenex \scriptfont3=\tenex \scriptscriptfont3=\tenex
\textfont\itfam=\eightit \def\it{\fam\itfam\eightit}
\textfont\slfam=\eightsl \def\sl{\fam\slfam\eightsl}
\textfont\ttfam=\eighttt \def\tt{\fam\ttfam\eighttt}
\textfont\bffam=\eightbf
\scriptfont\bffam=\sixbf
\scriptscriptfont\bffam=\fivebf \def\bf{\fam\bffam\eightbf}
\tt \ttglue=.5em plus.25em minus.15em
\normalbaselineskip=6pt
\setbox\strutbox=\hbox{\vrule height7pt width0pt depth2pt}
\let\sc=\sixrm \let\big=\eightbig \normalbaselines\rm}
\newinsert\footins
\def\newfoot#1{\let\@sf\empty
  \ifhmode\edef\@sf{\spacefactor\the\spacefactor}\fi
  #1\@sf\vfootnote{#1}}
\def\vfootnote#1{\insert\footins\bgroup\eightpoint
  \interlinepenalty\interfootnotelinepenalty
  \splittopskip\ht\strutbox 
  \splitmaxdepth\dp\strutbox \floatingpenalty\@MM
  \leftskip\z@skip \rightskip\z@skip
  \textindent{#1}\footstrut\futurelet\next\fo@t}
\def\fo@t{\ifcat\bgroup\noexpand\next \let\next\f@@t
  \else\let\next\f@t\fi \next}
\def\f@@t{\bgroup\aftergroup\@foot\let\next}
\def\f@t#1{#1\@foot}
\def\@foot{\strut\egroup}
\def\footstrut{\vbox to\splittopskip{}}
\skip\footins=\bigskipamount 
\count\footins=1000 
\dimen\footins=8in 

\def\ref#1{$^{#1}$}
\def\flex{\raise 6pt\hbox{$\leftrightarrow $}\! \! \! \! \! \! }
\def\oversome#1{ \raise 8pt\hbox{$\scriptscriptstyle #1$}\! \! \! \! \! \! }
\def\tr{ \mathop{\rm tr}}

\newbox\bigstrutbox
\setbox\bigstrutbox=\hbox{\vrule height10pt depth5pt width0pt}
\def\bigstrut{\relax\ifmmode\copy\bigstrutbox\else\unhcopy\bigstrutbox\fi}
\def\refer[#1/#2]{ \item{#1} {{#2}} }
\def\rev<#1/#2/#3/#4>{{\it #1\/} {\bf#2}, {#3}({#4})}
\def\boxit#1{\vbox{\hrule\hbox{\vrule\kern3pt
\vbox{\kern3pt#1\kern3pt}\kern3pt\vrule}\hrule}}

\def\2figure#1#2#3#4{\vbox{ \hrule width#1truecm \hbox{\vrule height#2truecm
\hskip #1truecm
\vrule height#2truecm }\hrule width#1truecm \hbox{\vrule\vbox{\hsize #1truecm
\baselineskip=10pt
\noindent\strut#3}\vrule}\hrule width#1truecm
\hbox{\vrule\vbox{\hsize #1truecm
\baselineskip=10pt
\noindent\strut#4}\vrule}\hrule width#1truecm  }}
\def\3figure#1#2#3#4#5{\vbox{ \hrule width#1truecm \hbox{\vrule height#2truecm
\hskip #1truecm
\vrule height#2truecm }\hrule width#1truecm \hbox{\vrule\vbox{\hsize #1truecm
\baselineskip=10pt
\noindent\strut#3}\vrule}\hrule width#1truecm
 \hbox{\vrule\vbox{\hsize #1truecm
\baselineskip=10pt
\noindent\strut#4}\vrule}
\hrule width#1truecm \hbox{\vrule\vbox{\hsize #1truecm
\baselineskip=10pt
\noindent\strut#5}\vrule}\hrule width#1truecm  }}

\def\sqr#1#2{{\vcenter{\hrule height.#2pt
   \hbox{\vrule width.#2pt height#1pt \kern#1pt
    \vrule width.#2pt}
    \hrule height.#2pt}}}


\def\smin{\,\raise 0.06em \hbox{${\scriptstyle \in}$}\,}
\def\smsubset{\,\raise 0.06em \hbox{${\scriptstyle \subset}$}\,}

\def\Natural{\hbox{\hskip 1.5pt\hbox to 0pt{\hskip -2pt I\hss}N}}

\def\Rational{\hbox{\hbox to 0pt{\hskip 2.7pt \vrule height 6.5pt
                                  depth -0.2pt width 0.8pt \hss}Q}}
\def\Real{\hbox{\hskip 1.5pt\hbox to 0pt{\hskip -2pt I\hss}R}}
\def\Complex{\hbox{\hbox to 0pt{\hskip 2.7pt \vrule height 6.5pt
                                  depth -0.2pt width 0.8pt \hss}C}}

\def \E {{{\rm e}}}


\def \1ok{{1\over \kappa ^2} }

\def \3dslim {{\rm DS}\!\!\!\!\!\!\!\!\lim }
\def \4dslim {{\rm DS}\!\!\!\!\!\!\!\!\!\!\lim }
\def \tr {{\rm tr}\, }
\def \ln {{\rm ln}\, }
\def \2kk{\left( \matrix {2k\cr k\cr }\right) }
\def \Rs4{{R^k\over 4^k} }

\def \1ok{{1\over \kappa ^2} }



\def \E {{\rm e}}
\nopagenumbers
\hfill CERN-TH-7354/94

\hfill hepth 9407128

\centerline{\grrm Integrability and duality in two-dimensional QCD}
\vskip 1.5cm
\centerline {E. Abdalla\newfoot {${}^*$}{Permanent address: Instituto de
F\'\i sica, USP, C.P. 20516, S. Paulo, Brazil.} and  M.C.B. Abdalla\newfoot
{${}^{**} $}{Permanent address: Instituto de F\'\i sica Te\'orica - UNESP,
R. Pamplona 145, 01405-000, S.Paulo, Brazil.}, }
\vskip 1.5truecm
\centerline { CERN, Theory Division, CH-1211 Geneva 23, Switzerland}
\vskip 2cm
\centerline{\bf Abstract}
\vskip .5cm
\noindent We consider bosonized QCD$_2$, and prove that after rewriting the
theory in terms of gauge-invariant fields, there exists an integrability
condition that is valid for the quantum theory as well. Furthermore,
performing a
duality-type transformation we obtain an appropriate action for the
description of the strong coupling limit, which is still integrable. We also
prove that the model displays a complicated set of constraints, restricting
the dynamics of part of the theory, but which are necessary to maintain the
positive metric Hilbert space.
\vfill

\noindent CERN-TH.7354/94

\noindent July 1994

\noindent hep-th 9407128
\vskip 1cm
\eject
\countdef\pageno=0 \pageno=1
\newtoks\footline \footline={\hss\tenrm\folio\hss}
\def\folio{\ifnum\pageno<0 \romannumeral-\pageno \else\number\pageno \fi}
\def\advancepageno{\ifnum\pageno<0 \global\advance\pageno by -1
\else\global\advance\pageno by 1 \fi}

\centerline {\bf 1. Introduction}
\vskip .5cm
\noindent In contrast with the case of the Schwinger model, quantum
chromodynamics of
massless fermions in 1+1 dimensions cannot be solved in terms of free fields.
Several methods have been used in such a case, some of them giving useful
results. We mention here the $1/N$ expansion introduced by 't Hooft\ref{1},
from which one obtains some information about the spectrum of the theory, and
the computation of the exact fermion determinant\ref{2} in terms of a Wess--\-
Zumino--Witten model\ref{3,4}, by which one arrives at an equivalent bosonic
action\ref{2,5}. Several authors made efforts in the direction of solving such
a difficult model\ref{6}, but an exact solution is still missing (see [7] for
an extensive review).

Working in the light cone gauge $(A_-=0)$ and formulating the problem in terms
of light cone variables, 't Hooft obtained a non-linear equation for the
fermion
self energy, from which he could obtain the above-mentioned information about
the spectrum of the theory. This procedure is however ambiguous, as pointed out
by Wu\ref{8}, and implies a tachyon for small bare fermion masses (therefore
also in the massless fermions case), see also [9]. This situation clearly
requires that a non-perturbative and explicitly gauge-invariant approach should
be used in order to obtain information based on firm grounds. Some authors
speculated that this situation was a sign of a possible non-trivial phase
structure of the theory\ref{26}.

The model has also been extensively studied, especially in the absence of
fermions, in relation to string theory. Indeed, string theory should be a model
describing bound states in strong interactions. Therefore Wilson loops should
there be described by the string approach as well\ref{10}. In this sense it is
natural to integrate out the fermions obtaining gauge-invariant objects, which
describe mesonic bound states with an infinite string attached to it. In such a
case, the construction of fields such as $\Sigma \!=\! UV$ (see sec. 2) or
$\widetilde g \!= \!UgV$, describing gauge-invariant bound states, is natural
after all. The original content of the (gauge-dependent) fermion fields can
still be recovered from the source terms. Those are kept as a bookkeeping
concerning the translation between original gauge-\-dependent fields and the
bosonic formulation, which is non-local. In this sense, we also recall the
construction of bilinear fermion fields\ref{11} in the light cone gauge, which
arises also from a WZW-type action.

The fact that QCD$_2$ can be studied using non-perturbative methods is by
itself a non-trivial statement. In the light cone gauge this can be motivated
by the fact that bilinears in the fermion fields form a $W_\infty$
algebra\ref{11}, which underlies integrable theories\ref{12}. Here, we shall
see that one obtains higher-conservation laws, building an affine Lie algebra.
We use methods based on the Polyakov--Wiegman identity\ref{2} extensively used
in refs. [13] and [14], and the consequent duality transformations\ref{14},
which seem to have a widespread application in such a class of
models\ref{14,15}. Due to the underlying gauge invariance, imposition of the
BRST condition in order to obtain the spectrum will be of crucial
importance\ref{14,16}.

Having in mind the above motivations, we first rewrite in section 2 the
problem in terms of bosonic matrix variables. Fermionic Green functions can be
obtained from the sources, which have been kept in the process. However, the
path integration is performed in terms of bosonic fields. We found a set of
fields in terms of which the partition function factorizes, as a product of two
conformal theories, and a third model corresponding to an off-critical
perturbation of the WZW theory. Later we verify that the different sectors
interact non-trivially due to the constraint structure. However we proceed
with a semi-\-classical reasoning, computing the equation of motion of the
latter
field, proving that it corresponds to an integrability condition of the theory.
Notice that due to the fermionic integration, there are already, at this point,
corrections of the order of $\hbar$. We verify that there is a non-trivial
change of variables that leads to a dual formulation of the theory, in terms
of the fields appropriatly to describing the strong coupling limit. Using
the Poisson bracket structure in section 4, we verify that the higher
conservation laws obey a Kac--Moody algebra. We also argue that the
quantization
of such higher conservation laws can be done by means of the introduction of
renormalization factors for the current contribution. In section 5 we discuss
the constraints arising from the structure of the gauge interaction, and
subsequently (section 6) we abtain also second class constraints. In section 7
we discuss the consequence of the constraints for the dual theory, and later
we argue about the Regge behaviour of the spectrum\ref{1}. We still discuss
the possibility of a conformally-invariant type solution for the current,
ending with some further conclusions.

\vskip 2cm
\penalty-1000
\centerline {\bf 2. Bosonization of two-dimensional QCD}
\vskip .5cm
\nobreak
\noindent We shall consider the QCD$_2$ Lagrangian, given by the expression
$$
{\cal L} = -{1\over 4}\tr F_{\mu\nu}F_{\mu\nu} + \overline \psi i \not \! \! D
\psi \quad ,\eqno(2.1)
$$
where $D_\mu = \partial _\mu - ie A_\mu$. We work out  some results in the path
integral formulation, but at a later stage consider also the canonical
quantization, forcing us to use both Euclidean and Minkowski spaces
alternatively. The conventions are given in the appendix.

The partition function is
$$
{\cal Z} \left[ \overline \eta , \eta , i_\mu \right] = \int {\cal D}\psi
{\cal D}\overline \psi {\cal D}A_\mu \, \E ^{-\int {\rm d}^2 z\,{\cal L}  -
\int
{\rm d}^2 z \, (\overline \eta  \psi + \overline \psi \eta  + i_\mu A_\mu ) }
\quad ,\eqno(2.2)
$$
where $\eta , \overline \eta $ are the external sources for the fermions
$\overline \psi, \psi$, and $i_\mu$ the external source for the gauge field
$A_\mu$.

In order to obtain the bosonized version of the theory one has  to rewrite the
fermionic determinant $\det i\not\!\!\! D$ as a bosonic functional integral. To
achieve this we consider the change of variables\ref{2}
$$
\eqalign{
\overline A & = {i\over e} V\overline \partial V^{-1}\quad ,\cr
 A & = {i\over e} U^{-1} \partial U\quad .\cr}\eqno(2.3)
$$

The fermionic determinant is given up to factors of the free Dirac operator, to
be discussed later in connection with the ghost system, by the expression
$$
\det i \not \!\! D = \E^{\Gamma[UV]} \quad ,\eqno(2.4)
$$
where the $\Gamma[g]$ is the Euclidean Wess--Zumino--Witten (WZW) functional,
given by the expression
$$
\Gamma[g] = {1\over 8\pi}\int {\rm d}^2 z\, \partial _\mu g^{-1} \partial _\mu
g
- {i\over 4\pi} \epsilon ^{\mu\nu} \int {\rm d} r \int {\rm d}^2z\,\hat g
^{-1} \dot {\hat g} \hat g^{-1} \partial _\mu \hat g
\hat g^{-1} \partial _\nu \hat g \quad ,\eqno(2.5)
$$
where $\hat g(r,z,\overline z)$ is the usual extension of $g(z\overline z)$ to
a space having the  Euclidean two-di\-mensional space as a boundary, and
$\hat g(1,z,\overline z)=g(z,\overline z)\,,\, \hat g(0,z,\overline z)=1$.
The WZW functional obeys the Polyakov--Wiegman identity\ref{2}
$$
\Gamma[UV] = \Gamma[U] + \Gamma[V]+{1\over 4\pi}\tr\int{\rm d}^2z \,U^{-1}
\partial U \,V\overline \partial V^{-1}\quad ,\eqno(2.6)
$$
which will be used extensively in this work. We find the seeked bosonized
formulation, taking advantage of the invariance of the Haar measure, and write
the above fermion determinant as
$$
\det i \not \!\! D \equiv \E^{-W(A)} = \int {\cal D} \, g \, \E^{-S_F[A,g]}
\quad ,\eqno(2.7a)
$$
where $S_F [A,g]$ is the equivalent of the fermionic action in terms of the
bosonic variables and gauge field and reads
$$
\eqalign{
S_F [A,g] & = \Gamma[UgV] - \Gamma[UV] \cr
& = \Gamma [g] + {1\over 4\pi} \int {\rm d}^2z\, \left[e^2 A_\mu A_\mu-e^2 A g
\overline A g^{-1} - ie A g \overline \partial g^{-1} - ie \overline A g^{-1}
\partial g \right] \quad .\cr}\eqno(2.7b)
$$
The above equation was obtained by a repeated use of the Polyakov--Wiegman
identity (2.6), respecting local gauge transformations. The external sources
have been used to redefine the integration over the fermionic field as
$$
\overline \psi i\not\!\!D \psi + \overline \eta  \psi + \overline \psi \eta
= \left( \overline\psi+\overline \eta (i\not\!\! D)^{-1}\right) i\not\!\! D
\left( \psi+(i\not\!\! D) ^{-1}\eta\right) - \overline \eta (i\not\!\! D)
^{-1} \eta  \quad .\eqno(2.8)
$$

The change of variables (2.3) leads to a non-trivial Jacobian, but fortunately
it is also the exponential of the WZW functional, that is
$$
{\cal D}A {\cal D}\overline A = \E^{c_V \Gamma[UV]}{\cal D}U{\cal D}V\quad,
\eqno(2.9)
$$
where $c_V$ is the quadratic Casimir, and a definite regularization respecting
vector gauge invariance has been chosen. As we  stressed after (2.3), this is
again written up to a factor containing the free Dirac operator.

The non-linearity in the gauge field interaction can also be disentangled by
means of the identity (see notation in the appendix)
$$
\E^{{1\over 4}\int{\rm d}^2 z\,\tr F_{\mu\nu}^2 } = \int {\cal D}E\, \,\E^{\int
{\rm d}^2z\,\left[{1\over 2} \tr E^2 + {1\over 2}\tr E F_{z\bar z}\right]}
\quad , \eqno(2.10)
$$
where $E$ is a matrix-valued field. Taking into account the previous set of
information we arrive at
$$
\eqalign{
&{\cal Z}\left[ \overline \eta , \eta , i_\mu\right]= \int {\cal D}E{\cal D}
U{\cal D}V{\cal D}g \times \cr
& \times \E^{-\Gamma[UgV] + (c_V+1) \Gamma[UV] + \int {\rm d}^2 z \,\tr
[{1\over 2} E^2 + {1\over 2}E F_{z\bar z}] - \int {\rm d}^2z\, i_\mu A_\mu
+ \int {\rm d}^2z {\rm d}^2w \, \overline \eta  (z) (i \not D)^{-1}(z,w)
\eta (w) }\quad .\cr}\eqno(2.11)
$$

If we were considering massive fermions, we should include a term $\, m\tr (g+
g^{-1})$ in the effective action\ref{6}. However we shall avoid such a
complication
and only consider the massless case. We also have to deal with gauge fixing. In
fact, the process of introducing ghosts is standard, and we suppose that the
procedure is included above, until it is necessary to explicitly take into
account the ghost degrees of freedom, which will be the case upon consideration
of the spectrum, when the BRST condition has to be used. Up to that point our
manipulations do not explicitly depend on the gauge fixing/ghost system, and
we proceed without it (or else, keeping it at the back of our minds).

Defining the gauge invariant field $\widetilde g= UgV$, and using the
invariance of the Haar measure, that is ${\cal D} g = {\cal  D} \widetilde g$,
we see that the $\widetilde g$ field decouples (always up to BRST condition,
see later) and we are left with
$$
\eqalign{
&{\cal Z}\left[ \overline \eta , \eta , i_\mu\right]= \int {\cal D}
\widetilde g \, \E^{-\Gamma[\tilde g]}\int  {\cal D}E {\cal D}U {\cal D}V
{\cal D} ({\rm ghosts})\times \cr
& \times \E^{(c_V+1)\Gamma[UV] + \int{\rm d}^2z\, \tr[{1\over 2} E^2 +
{1\over 2}EF_{z\bar z}] - S_{ghosts} - \int {\rm d}^2 z\, i_\mu A_\mu +
\int {\rm d}^2z {\rm d}^2w \, \overline \eta  (z) (i \not D)^{-1}(z,w)
\eta  (w) }\quad ,\cr}\eqno(2.12)
$$
where $A_\mu$ variables are given in terms of the $UV$ variables.

The presence of the gauge field strength $F_{z\overline z}$ hinders further
developments in the way it is presented above. However, in terms of the $U$
and $V$ variables we can write
$$
\tr E F_{z\bar z}= {i\over e} \tr UEU^{-1} \partial (\Sigma \overline
\partial\Sigma ^{-1}) \quad ,\eqno(2.13)
$$
where $\Sigma = UV$. This will permit a complete separation of some variables.
Indeed, $\Sigma$ is a more natural candidate to represent the physical degrees
of freedom, since $U$ and $V$ are not separately gauge-invariant. In the way
it is written in eq. (2.13), we can redefine $E$ taking advantage once more of
the invariance of the Haar measure, in such a way that the effective action
only
depends on the combination $\Sigma = UV$, while $U$ and $V$ appear separately
only in the source terms, which are gauge-dependent, as they should be, that
is,
$$
\eqalign{
A& = {i\over e} U^{-1}\partial U\quad ,\cr
\overline A&={i\over e}(U^{-1}\Sigma)\overline\partial(\Sigma^{-1} U)\quad
.\cr}
\eqno(2.14)
$$
If we eventually choose the light cone gauge, $U=1,\,\overline A={i\over e}
\Sigma\overline\partial\Sigma^{-1}$ and $A=0$. From the structure of (2.13), it
is natural to redefine variables as $\widetilde E'=UEU^{-1},{\cal D} E={
\cal D}\widetilde E'$. Notice that already at this point the $E$ redefinition
implies, in terms of the gauge potential, an infinite gauge tail, which
captures the possible gauge transformations. It is also convenient to make
the rescaling $\widetilde E'=2ie(c_V+1)\widetilde E$, with a constant Jacobian.
 In terms of the field $\widetilde E$, consider the change of variables
$$
\partial \widetilde E = {1\over 4\pi} \beta^{-1}\partial \beta \quad
,\quad {\cal D}\widetilde E = \E^{c_V\Gamma[\beta]}{\cal D} \beta\quad .
\eqno(2.15)
$$
We use the identity (2.6) to transform the $\beta \Sigma$ interaction into
terms that can be handled in a more appropriate fashion. Writing both steps
separately we have
$$
\eqalign{
&{\cal Z}\left[\overline \eta , \eta , i_\mu\right]=\int{\cal D}\widetilde g
\, \E^{-\Gamma[\tilde g]}  {\cal D}U  {\cal D} ({\rm ghosts}) \, \E^{- S_{{\rm
ghosts}}}
\int {\cal D} \Sigma {\cal D} \widetilde E \E^{(c_V+1)\Gamma[\Sigma]}\times \cr
& \!\times \!\E^{(c_V+1)\tr\!\int\!{\rm d}^2z\partial\widetilde E\Sigma
\overline\partial\Sigma^{^{-1}}\!
-\!2e^2 (c_V+1)^2 \!\int \!{\rm d}^2z\tr \widetilde E^2 - \!\int \!{\rm d}^2
z \,i_\mu A_\mu + \int {\rm d}^2z {\rm d}^2w \,\overline \eta
(z)(i\not D)^{^{-1}}(z,w)\eta (w)}\cr}\eqno(2.16)
$$
in such a way that after substitution of (2.15) into (2.16) and using (2.6)
for $\Gamma [\beta \Sigma]$, we arrive at
$$
\eqalign{
&{\cal Z}\left[\overline \eta , \eta , i_\mu\right]= \int{\cal D}\widetilde g
\, \E^{-\Gamma[\tilde g]}  {\cal D} U  {\cal D}({\rm ghosts})\, \E^{-
S_{{\rm ghosts}}} \int {\cal D} \Sigma {\cal D} \beta \times \cr
&\!\times \!\E^{(c_V+1)\Gamma[\beta\Sigma]
-\Gamma[\beta]- {2e^2(c_V+1)^2\over (4\pi)^2}\tr\int\!{\rm d}^{^2}z[\partial^
{^{-1}}(\beta^{^{-1}}\partial\beta)]^2-\int\!{\rm d}^{^2} z i_\mu A_\mu + \int
{\rm d}^{^2}z {\rm d}^{^2}w \overline \eta (z) (i\not D)^{^{-1}}\!(z,w)
\eta (w)} .\cr}\eqno(2.17)
$$

We define the (massive) parameter
$$
\lambda = {c_V+1 \over 2\pi}e \quad ,\eqno(2.18)
$$
and the field $\widetilde \Sigma = \beta\Sigma$, in terms of which the
partition function reads
$$
\eqalign{
&{\cal Z}\left[ \overline \eta , \eta , i_\mu\right]=\int{\cal D}\widetilde g
\, \E^{-\Gamma[\tilde g]}  {\cal D} U  {\cal D} ({\rm ghosts})\, \E^{-
S_{{\rm ghosts}}}
\int {\cal D} \widetilde \Sigma
\,\E^{(c_V+1)\Gamma[\widetilde\Sigma]}\times\cr
&\times \int {\cal D} \beta\, \E^{-\Gamma[\beta]- {\lambda^2\over 2}\tr \int
{\rm d}^2 z\,[\partial^{-1}(\beta^{-1}\partial\beta)]^2}\,\E^{-\int{\rm d}^2z\,
i_\mu A_\mu +\int{\rm d}^2z\,{\rm d}^2w\overline \eta (z) (i \not D)^{-1}(z,w)
\eta (w) }\quad ,\cr}\eqno(2.19)
$$
where now $A = {i\over e}U^{-1}\partial U\, ,\, \overline A = {i\over e}
(U^{-1}\beta^{-1}\widetilde \Sigma )\overline\partial(\widetilde\Sigma^{-1}
\beta U)$.

Up to source terms, and the BRST constraints to be discussed later, the above
generating functional factorizes in terms of a conformal theory for $\widetilde
g$, representing a gauge-invariant fermionic bound states degrees of
freedom, a second conformal field theory for $\widetilde \Sigma$, representing
some gauge condensate, and an off-critically perturbed conformal field theory
for the $\beta$-field, which describes also a gauge field condensate, in view
of the change of variables (2.15).  The conformal field theory representing
$\widetilde \Sigma$ has an action with a negative sign (see (2.19)). Therefore
we have to carefully take into account the BRST constraints in order to arrive
at a positive metric Hilbert space. Since we will study first the
$\beta$-degrees of freedom, we leave this problem for a later section.

\vskip 2cm
\penalty-3000
\centerline {\bf 3. Integrable perturbation of the WZW theory and duality}
\vskip .5cm
\nobreak
\noindent We consider the perturbed WZW action
$$
\eqalign{
S &= \Gamma[\beta] + {1\over 2} \lambda^2 \tr \int {\rm d}^2z\, \left[\partial^
{-1}(\beta^{-1} \partial \beta)\right] ^2\quad ,\cr
&= \Gamma[\beta] + {1\over 2} \lambda^2 \Delta(\beta)\quad .\cr}\eqno(3.1)
$$

We will look for the Euler--Lagrange equations for $\beta$. It is not difficult
to find the variations:
$$
\eqalign{
\delta \Gamma[\beta ]&= \left[ -{1\over 4\pi} \overline \partial (\beta^{-1}
\partial \beta)\right]\beta^{-1}\delta\beta\quad ,\cr
\delta \Delta(\beta) & = 2\Big[ \partial ^{-1} (\beta^{-1}\partial \beta)
 + \big[ \partial ^{-2} (\beta^{-1}\partial \beta), (\beta^{-1}\partial\beta)
\big] \Big]\beta^{-1}\delta \beta\quad .\cr} \eqno(3.2)
$$

Collecting the terms, we find it useful to define the current components
$$
\eqalign{
J^\beta&= \beta^{-1}\partial \beta\quad ,\cr
\overline J^\beta&= 4\pi \lambda^2 \partial ^{-2}J^\beta = 4\pi \lambda^{2}
\partial ^{-2}(\beta^{-1}\partial \beta)\quad ,\cr}\eqno(3.3)
$$
which summarize the $\beta $ equation of motion as a zero curvature condition
given by
$$
[{\cal D}, \overline {\cal D}] = [\partial - J^\beta, \overline\partial -
\overline J^\beta]= \overline \partial J^\beta - \partial \overline J^\beta -
[\overline J^\beta, J^\beta ] =0 \quad .\eqno(3.4)
$$
This is the integrability condition for the Lax pair\ref{17}
$$
{\cal D}_\mu M =0 \quad  , \quad {\rm with } \quad {\cal D}_\mu = \partial _
\mu - J^\beta_\mu \quad ,\eqno(3.5)
$$
where $J^\beta=J^\beta_1+iJ^\beta_2\, ,\, \overline J^\beta= J^\beta_1 -
iJ^\beta_2$ and $M$ is the monodromy matrix. This is not a Lax pair as in usual
non-linear sigma models\ref{18}, where $J^\beta_\mu$ is a conserved current,
and we obtain a conserved non-local charge from (3.4), as well as
higher local and non-local conservation laws, derived from an extension of
(3.4) in terms of an arbitrary spectral parameter\ref{18}. However, to
a certain extent, the situation is simpler in the present case, due to the
rather unusual form of the currents (3.3), which permits us to write the
commutator appearing in (3.4) as a total derivative, in such a way that in
terms of the current $\overline J^\beta$, we have
$$
\partial \left\{ 4\pi \lambda^2 \overline J^\beta -\partial\overline\partial \,
\overline J^\beta + [\overline J^\beta, \partial \overline J^\beta]\right\} =0
\quad .\eqno(3.6)
$$

Therefore the quantity
$$
\overline I^\beta (\overline z) = 4\pi \lambda^2 \overline J^\beta (z,
\overline z) - \partial \overline \partial \overline J^\beta (z,\overline z) +
[\overline J^\beta (z, \overline z), \partial \overline J^\beta(z,\overline z)]
 \quad \eqno(3.7)
$$
does not depend on $z$, and it is a simple matter to derive an infinite number
of conservation laws from the above (see later in the Minkowskian formulation).

This means that two-dimensional QCD is an integrable system! Moreover, it
corresponds to an off-critical perturbation of the WZW action. If we write
$\beta=\E^{i\phi} \sim 1+ i\phi$,  we verify that the perturbing term
corresponds to a mass term for $\phi$. Later we will discuss in more detail
this
issue in the large-$N$ limit (for the $SU(N)$ theory). The next natural step is
to obtain the algebra obeyed by (3.7), and its representation. However, there
is
a difficulty presented by the non-locality of the perturbation. We now
introduce a further auxiliary field defining a dual action, local in all
fields, and representing the low-energy scales of the theory, and later we
return to the problem of finding the algebra obeyed by (3.7).

Consider the $\Delta$-term of the action (3.1). We write the quadratic term in
(3.1), introducing the integral over a Gaussian field $\partial \overline C$ as
$$
\E^{-{1\over 2}\lambda^2\Delta} = \int {\cal D} \overline C \, \E^{ \int
{\rm d}^2 z\,{1\over 2} \tr (\partial \overline C)^2 - \lambda \tr \int {\rm d}
^2z\,\overline C (\beta^{-1} \partial \beta)}\quad ,\eqno(3.8)
$$
where the left-hand side is readily obtained by completing the square in the
r.h.s.

Indeed, at this point we have two choices. We can turn to Minkowski space, and
proceed with the canonical quantization of the action (3.1) with the non-local
term substituted in terms of the $\overline C$ field-dependent expression
obtained in the exponent of the integrand of the r.h.s. of eq. (3.8).
Before that, motivated by the presence of the auxiliary vector field $\overline
C$, we make again a change of variables of the type
$$
\eqalign{
\overline C &= {1\over 4\pi \lambda} W\overline \partial W^{-1} \quad ,\cr
{\cal D} \overline C & = \E^{c_V \Gamma [W]} {\cal D} W \quad ,\cr}
\eqno(3.9)
$$
together with the now very frequently used identity (2.6) in order to find a
dual action. We have for the $\beta$-partition function the expression
$$
{\cal Z} = \int {\cal D}\beta {\cal D}W \,\E^{-\Gamma[\beta] + c_V \Gamma[W] -
{1\over 4\pi} \int {\rm d}^2z\, W\overline \partial  W^{-1} \beta^{-1} \partial
\beta + \int {\rm d}^2 z \,{1\over 2(4\pi \lambda)^2} [\partial (W\overline
\partial W^{-1})]^2}\quad ,\eqno(3.10)
$$
from which we can separate the contribution $-\Gamma[\beta W] \equiv -
\Gamma[\widetilde \beta]$; after such man{\oe}uvre we are left with
$$
{\cal Z} = \int {\cal D} \widetilde \beta \,\E^{-\Gamma[\widetilde \beta]}\int
{\cal D} W \,\E^{ (c_V+1)\Gamma[W]+{\tr\over 2(4\pi\lambda)^2} \int {\rm d}^2 z
\, [\partial(W\overline \partial W^{-1})]^2}\quad .\eqno(3.11)
$$

The dual action has now a coupling constant corresponding to the inverse of the
initial charge. Therefore (3.11) is appropriate to the study of a strongly
coupled limit. Notice that the procedure is, in a sense, familiar to the one
used to obtain a dual action, where a non-dynamical field is introduced, and
one
eliminates the original fields by integration leaving the so-called dual
formulation. See refs. [13], [14] and [15] for further details. We
separate a further WZW-conformal piece, and we are left with a local massive
action for $W$. The drawback is the fact that now $W$ itself has an action
with a negative sign. Na\"{\i}vely it describes also massive excitations,
although a complete description of the spectrum can only be obtained after
disentangling the non-linear relations and imposing the BRST conditions.

For the sources, we now have to replace $A$ in (2.19) by
${i\over e}(U^{^{-1}} W \widetilde\beta^{^{-1}} \widetilde \Sigma )\overline
\partial(\widetilde \Sigma^{^{-1}}\widetilde \beta  W^{^{-1}}  U)$. We notice
also here that we have dual descriptions of two-dimensional QCD. In the
first, valid in the perturbative region, for high energies, we find out a
non-local perturbation of the WZW action. In terms of $W$ the perturbation is
local, but at the price of a negative sign in the na\"{\i}ve kinetic term in
the $W$ action, which is appropriate to describing the low energy (strong
coupling) regime of the theory. In spite of such different complementary
descriptions, both models are integrable. In the weak coupling regime we found
the conservation laws (3.3 -- 3.6). In the case of the $W$-theory, it is not
difficult to find the equations of motion, and again derive  similar
relations for the quantity
$$
\overline I^W (\overline z)={1\over 4\pi}(c_V+1)\overline J^W (z,\overline z) +
{1\over (4\pi\lambda)^2}\partial\overline\partial\,\overline J^W(z,\overline z)
+ {1\over (4\pi\lambda)^2} [\overline J^W(z,\overline z), \partial\overline
J^W](z,\overline z)\quad ,\eqno(3.12)
$$
with $\overline J^W = W\overline \partial W^{-1}$ and $\partial\overline
I^W=0$, i.e. $\overline I^W$ does not depend on $z$.

Therefore, after finding isomorphic higher charges for both formulations, we
are motivated to find their corresponding algebras, and later quantize them.
\vskip 2cm
\penalty-300
\centerline {\bf 4. Higher conservation laws  and corresponding algebras }
\vskip .5cm
\nobreak
\noindent To obtain the algebra obeyed by the previously found conserved
charges, it is easier to go to Minkowski space, proceed with the canonical
quantization\ref{19}, obtaining first the Poisson algebra, and later the
constraints and the quantum commutators of the model. In fact, from the
computation of the fermion determinant, we have an effective bosonic action
that already takes into account some quantum corrections, namely the fermionic
loops have been summed up. Therefore, the Poisson brackets already have
quantum corrections arising from fermionic loops. This fact minimizes the
possibilities of anomalies in the full quantum definition of the
charges\ref{20}. As a matter of fact, we shall see that quantum corrections
are restricted to the introduction of renormalization constants.

{}From the conventions described in the appendix we find the Minkowski space
action
$$
S = - (c_V+1) \Gamma_M[W] + {1\over 2(4\pi \lambda)^2} \int {\rm d}^2 x \,
\left[ \partial _+ (W\partial _-W^{-1})\right]^2 \quad ,\eqno(4.1)
$$
with the Minkowski space WZW functional given by
$$
\Gamma_M[W] = {1\over 8\pi} \tr\int{\rm d}^2x \,\partial ^\mu W^{-1} \partial
_\mu W + {1\over 4\pi} \epsilon ^ {\mu\nu} \tr \int _0^1 {\rm d}r\int{\rm d}^2
x \,\hat W^{-1}\dot {\hat W} \hat W^{-1}\partial _\mu
\hat W \hat W^{-1}\partial _\nu \hat W \quad .\eqno(4.2)
$$
Due to the presence of higher derivatives in the above action, it is
convenient to introduce an auxiliary field and rewrite it in the equivalent
form
$$
S= -(c_V+1) \Gamma_M[W] + \tr {1\over 2}\int{\rm d}^2x\,\left[-B^2+{1\over 2\pi
 \lambda } \partial _+ B \partial _-WW^{-1}\right]\quad ,\eqno(4.3)
$$
where (4.1) is obtained by completing the square in the $B$-term in (4.3). The
momentum canonically conjugated to the variable $W$ is
$$
\eqalign{
\Pi^W_{ij} &= {\partial S\over \partial \partial _0W_{ij}} = -{1\over 4\pi}
(c_V+1) \partial _0W^{-1}_{ji} - {1\over 4\pi} (c_V+1)A_{ji} + {1\over 4\pi
\lambda}(W^{-1} \partial _+B)_{ji}\quad ,\cr
& = \hat \Pi^W_{ij} - {1\over 4\pi} (c_V +1) A_{ji}\quad ,\cr}\eqno(4.4)
$$
where the first term is obtained from the principal sigma-model term in the WZW
action, the second arises from the pure WZW term, and the third one from the
interaction with the auxiliary field. It is convenient to separate the WZW
contribution $A_{ij}$ to the momentum, since the new variable $\widehat \Pi^W $
is local in the original fields. The treatment of the WZW term (second in the
right hand side above) follows closely the one introduced in [19], see also
[7]. An
explicit form for $A_{ij}$ cannot be obtained in terms of local fields, but we
only need its derivatives, which are not difficult to obtain, that is\ref{7,19}
$$
F_{ij;kl} = {\delta A_{ij}\over \delta W_{lk}} - {\delta A_{kl}\over
\delta W_{ji}} = \partial _1 W^{-1}_{il} W^{-1}_{kj} - W^{-1} _{il} \partial
_1 W^{-1}_{kj}\quad ,\eqno(4.5)
$$
in terms of which we have the Poisson bracket relation
$$
\eqalign{
\left\{ \hat \Pi^W_{ij}(x), \hat \Pi^W_{kl}(y)\right\} & = - {c_V +1\over
4\pi}\left( {\delta A_{lk} \over \delta W_{ij}} - {\delta A_{ji} \over \delta
W_{kl}}\right) \cr
&= {c_V+1 \over 4\pi} \left( \partial _1 W^{-1}_{jk} W^{-1}_{li} -
\partial _1 W^{-1} _{li} W^{-1}_{jk}\right)\delta(x^1-y^1)\quad .\cr}
$$
The momentum associated with the $B$ field is
$$
\Pi_{ij}^B = -{1\over 4\pi\lambda} (W\partial _-W^{-1})_{ji}\quad .\eqno(4.6)
$$
We can now list the relevant field operators appearing in the definition of the
conservation law (3.12), which we rewrite in Minkowski space as
$$
\eqalign{
I^W_-& = {1\over 4\pi} (c_V +1) J^W_- - {1\over (4\pi\lambda)^2} \partial_+
\partial_-J^W_- - {1\over (4\pi\lambda)^2} [ J^W_-, \partial_+J^W_-]\quad ,\cr
\partial _+  I^W_- &= 0 \quad .\cr}\eqno(4.7)
$$
In terms of phase-space variables, they are
$$
\eqalign{
J^W_- & = W\partial _-W^{-1} = - 4\pi \lambda \widetilde \Pi_B\quad ,\cr
\partial_+J^W_-&=-4\pi\lambda\partial_+\widetilde \Pi_B=4\pi\lambda B\quad ,\cr
\partial_+\partial_-J^W_-&= (4\pi\lambda)^2 [W\widetilde {\hat \Pi}^W - (c_V+1)
\lambda \widetilde \Pi_B] - (4\pi \lambda)\lambda (c_V+1) W'W^{-1}- 8 \pi
\lambda B'\quad ,\cr} \eqno(4.8)
$$
where the tilde means a transposition of the matrix indices. It is
straightforward to compute the Poisson algebra. We have
$$
\left\{I^W_{ij}(t,x), I^W_{kl}(t,y) \right\}  = \left[ I^W_{kj}
 \delta_{il} - I^W_{il}\delta_{kj}\right]
\delta(x^1\!-\!y^1)- \alpha \delta^{il}\delta^{kj} \delta'(x^1-y^1)\eqno(4.9)
$$
where $\alpha = {1\over 2\pi}(c_V+1)$. The current itself is a realization of
the Kac--Moody algebra, since
$$
\eqalign{
\left\{ I^W_{ij}(t,x), J^W_{-kl}(t,y) \right\} & = (J^W_{-kj}\delta_{il} -
J^W_{-il}\delta_{kj})\delta(x^1-y^1) + 2\delta_{il}\delta_{kj}\delta'(x^1-y^1)
\quad ,\cr
\left\{ J^W_{ij}(t,x), J^W_{-kl}(t,y^1) \right\} & = 0\quad .\cr}\eqno(4.10)
$$
We thus obtain a Kac--Moody algebra for $I_-^W$, and $J^W_-$ is a
representation of such an algebra, with a central extension. We shall return
to this discussion later, after consideration of the quantization of the
charge.

The Hamiltonian density can also be computed, and we arrive at the
phase-space expression
$$
\eqalign{
H_W= & \widetilde {\hat \Pi}^W W' + 4\pi \lambda\widetilde{\hat
\Pi}^W\widetilde
\Pi ^B W - \widetilde \Pi^BB' - 4\pi \lambda^2 (c_V +1)(\widetilde \Pi^B)^2\cr
& - 2(c_V+1)\lambda \widetilde \Pi ^B W'W^{-1} + {1\over 4\pi} (c_V+1)(W'
W^{-1})^2 +{1\over 2}B^2\quad ,\cr}\eqno(4.11)
$$
where $B'=\partial _1B\, ,\, W'=\partial _1W$; the above Hamiltonian can be
rewritten in a quadratic form in terms of the currents, although in such a
case we have also velocities, due to the appearance of the time derivatives:
$$
H_W = \alpha \left( J_1^W\right)^{2} - {1\over (4\pi \lambda)^2}\left[\partial
_+^2J^W_-J^W_+ +J^W_-\partial _-\partial _+J^W_- - (\partial _+J^W_-)^2\right]
\quad ,\eqno(4.12)
$$
where $J^W_1= {1\over 2}(J^W_+-J^W_-)$ and $J_+^W = W \partial _+W^{-1}$. At
this point we can compare the model with its $\beta$-formulation. In this case
we have the action
$$
S = \Gamma_M[\beta] + \lambda \tr \int {\rm d}^2 x\,C_- \beta ^{-1}\partial_+
\beta + {1\over 2} \tr \int {\rm d}^2x\, (\partial _+C_-)^2 \quad ,\eqno(4.13)
$$
where $C_-$ is the Minkowski space counterpart of $\overline C$ (see eq.
(3.8)).

The canonical quantization proceeds straightforwardly, and the relevant
phase-space expressions are obtained for $\overline J^\beta $ in (3.3), which
in Minkowski space, due to the $C_-$ equation of motion, reads
$$
\eqalignno{
J^\beta_- &= 4\pi \lambda^2 \partial _+^{-2} (\beta^{-1} \partial _+ \beta) =
4\pi \lambda C_-\quad ,&(4.14a)\cr
\Pi_- & = \partial _+ C_-\quad ,&(4.14b)\cr}
$$
while the $\beta$-momentum is given by
$$
\widetilde{\hat \Pi}^\beta_{ji}  = {1\over 4\pi} \partial _0 \beta^{-1}_{ji} +
\lambda (C_-\beta^{-1})_{ji}\quad ,\eqno(4.14c)
$$
where the hat above $\Pi^{{}^\beta}$ means that we neglected the WZW
contribution as before\ref{25}, and as a consequence
$$\left\{ \widetilde{\hat \Pi}^\beta_{ji}(t,x), \widetilde{\hat \Pi}^\beta_{lk}
(t,y)\right\} =-{1\over 4\pi}\left( \partial_1\beta^{{}^{-1}}_{_{jk}}
\beta^{{}^{-1}}_{_{li}}-\partial_1\beta^{{}^{-1}}_{_{li}}
\beta^{{}^{-1}}_{_{jk}}\right)\delta(x-y)\eqno(4.14d)$$
{}From the definition of the canonical momentum associated with $C_-$ we
have
$$
\partial _+ J^\beta_- = 4\pi \lambda \Pi _{-}\quad .\eqno(4.15)
$$
The conserved charge is (from (3.6) we change $\partial \to -\partial _-\, ,\,
\overline \partial \to \partial _+\, ,\, J^\beta_-\to J^\beta_-$):
$$
\eqalign{
I^\beta_- &= 4\pi \lambda^2J^\beta_- + \partial_+\partial_-J^\beta_-+[J^\beta_-
,\partial_+J^\beta_-]\quad,\cr
\partial _+I^\beta_-&=0\quad ;\cr} \eqno(4.16)
$$
therefore the situation is analogous to the one we found previously by
interchanging the $ (B, \Pi_B)$ phase-space variables with $(\Pi_-, C_-)$
(noticing the exchanged order).

At this point the Hamiltonian might be computed. However we will postpone it to
a later section, since we will have to compute it in terms of more appropriate
currents, making the problem easier to  formulate in terms of the
constraints that are hidden in the gauge transformation properties.

We now come to the point where we are urged to consider the quantization of the
symmetry current (4.16). Let us consider the problem in the $\beta$-language,
since the short-distance expansion depends on the high-energy behaviour of the
theory; therefore, since the only massive scale is the coupling constant, we
have to consider the weak coupling limit. The weak coupling limit is better
described by the $\beta$-action. In such a case, we need the short-distance
expansion of the current $J^\beta _-=4\pi\lambda^2\partial_+^{-2}(\beta^{-1}
\partial _+\beta)$ with itself. Since the short distance expansion is
compatible with the weak coupling limit, where the theory is conformally
invariant, Wilson expansions can be dealt with as usually.

We consider the short distance expansion
$$
[J^\beta_-(x), \partial _+J^\beta_-(y)]\Bigg\vert_{y=x+\epsilon} = \sum _n
a^{(n)}(\epsilon) {\cal O}^{(n)}(x)\quad ,\eqno(4.18)
$$
aiming at a classification of ${\cal O}^{(n)} (x)$ according to its own
dimension\ref{20}.
It is in fact easier to start out of local objects, that is in terms of
$J^\beta_+$:
$$
{ J^\beta_+ }={1\over 4\pi\lambda^2}\partial_+^2J^\beta_- = \beta^{-1}
\partial_+\beta\quad ,\eqno(4.19)
$$
and later act with antiderivative operators. Therefore we analyse the auxiliary
operator product expansion
$$
[{J^\beta_+} (x), {J^\beta_+} (y)]\Bigg\vert_{y=x+\epsilon} = \sum _n a_
{J^\beta_+ }^{(n)}(\epsilon) {\cal O}_{{J^\beta_+} }^{(n)}(x)\quad .\eqno(4.20)
$$

Let us suppose that the theory is local. The fact that the interaction contains
an antiderivative will be taken into account subsequently. Along such
premises the problem is very simple, and has been solved long ago\ref{20},
with the
classification for possible operators ${\cal O} _{{J^\beta_+} }^{(n)}(x)$:
$$
\eqalign{
1. \quad {\rm dim } \,{\cal O}^{(n)} =0 \quad &, \quad {\rm no \quad operator}
\quad , \cr
2. \quad {\rm dim } \, {\cal O}^{(n)} =1\quad &,\quad{\cal O}_{{J^\beta_+} }
^{(1)}(x)=\beta^{-1}\partial \beta \quad , \cr
3. \quad {\rm dim } \, {\cal O}^{(n)} =2\quad &,\quad{\cal O}_{{J^\beta_+}}^
{(2)}(x)= \partial(\beta^{-1}\partial \beta) \quad ,\cr
4. \quad {\rm dim } \, {\cal O}^{(n)} \ge 3\quad &, \quad {\rm operators
\quad with \quad finite \quad coefficients}\quad . \cr}
$$

Therefore we find
$$
[{J^\beta_+} (x), {J^\beta_+} (y)] = a^{(1)}(\epsilon) {J^\beta_+} (x) +
a^{(2)}
(\epsilon) \partial {J^\beta_+} (x)\quad ,\eqno(4.21)
$$
where ${\rm dim}\, a^{(1)}(\epsilon) = 1$, therefore $a^{(1)} (\epsilon)$ is
linearly divergent, that is, $a^{(1)} (\epsilon)\sim 1/\epsilon$, while
${\rm dim}\,  a^{(2)}=0$, and $a^{(2)} (\epsilon )$ is logarithmically
divergent, that is $a^{(2)} (\epsilon) \sim \ln \epsilon$. Acting on the above
Wilson expansion with $\partial _x^{-2}$, we obtain the new expansion
$$
[\partial _+^{-2} {J^\beta_+} (x), {J^\beta_+} (y)] = a^{(1)} (\epsilon)
\partial _+^{-2} {J^\beta_+} (x) + \widetilde a^{(2)} (\epsilon) \partial_+
^{-1} {J^\beta_+} (x)\quad ,\eqno(4.22)
$$
where $\widetilde a^{(2)} (\epsilon)$ is $a^{(2)} (\epsilon)$ plus a possible
$\partial^{-1}a^{(1)}(\epsilon)$ correction. We now act with $\partial _
\epsilon^{-1}$, obtaining
$$
[\partial _+^{-2} {J^\beta_+} (x), \partial _+^{-1} {J^\beta_+} (y)] =
\partial ^{-1} \widetilde a^{(2)} (\epsilon) \partial _+^{-2} {J^\beta_+} (x)
+ {\rm finite }\quad ,\eqno(4.23)
$$
since $\partial ^{-1}a^{(2)}$ is finite.

Let us now discuss the effect of the non-local term in the action. Its local
version is given by an expression containing the $C_-$ field (see [19]). Such a
field has dimension zero, and its interaction contains a $\lambda$ factor in
the
Lagrangian, namely
$$
{\cal L}_{C_-} ={1\over 2}(\partial_+ C_-)^2+ \lambda C_-\beta^{-1} \partial
_+\beta\quad .\eqno(4.24)
$$

Therefore $C_-$ comes in a Wilson expansion accompanied by a
$\lambda$-factor, and has, effectively, dimension 1. Moreover, it cannot
appear alone, since by Lorentz transformation it acquires a factor that is
the inverse of the one required for the current, since it is the $(-)$
component of a vector, while the current is a $(+)$ component. Therefore it can
appear at most with a logarithmically divergent coefficient in the ${J^\beta_+}
(x){J^\beta_+} (x+\epsilon)$ expansion, and is irrelevant to the present
problem, due to the subsequent manipulations.

This discussion leads to the definition of the cut-off charges
$$
Q_\delta^f = \int {\rm d} x^-\, f(x^-)\left\{ {\cal Z}_\delta J^\beta _-(x) +
[J^\beta _-(x), \partial J^\beta _-(x+\delta)] \right\} \quad ,\eqno(4.25)
$$
with renormalized charge $Q^f$ and renormalization constant ${\cal Z}_\delta$,
respectively given by
$$
Q^f = \lim _{\delta \to 0} Q_\delta ^f \quad {\rm and}\quad {\cal Z} _\delta =
1 -\partial ^{-1} a^{(1)}(\delta)\quad . \eqno(4.26)
$$
Such a charge is finite, and
$$
{d Q_\delta^f\over dx^+} = \lim _{\delta \to 0}\int {\rm d} x^{-}\, f(x^-)
\Big\{ {\cal Z} _\delta \partial _+J^\beta_- + \partial _+ \big[\partial ^{-1}
a^{(1)}J^\beta_- + {\cal N}[J^\beta_-,\partial_+ J^\beta_-]\big]\Big\}
\longrightarrow 0 \quad ,\eqno(4.27)
$$
where ${\cal N}$ is a normal product prescription rendering the product in
$[J^\beta _-, \partial_+ J^\beta _-]$ finite.

The infinite constant can also be interpreted as a charge renormalization. Due
to the renormalization of the higher charge, we cannot give an interpretation
to the field operator $I^\beta_{ij}$ by itself, but only to an arbitrary linear
combination involving the charge and the current. In  any case, since $I^
\beta_{ij}$ is a right-moving field operator, it is natural to assume, in view
of the Poisson algebra (4.9), that it obeys an algebra
given by\ref{21,22}
$$
I^\beta_{ij}(\overline z) I^\beta_{kl}(\overline w) = (I^\beta_{kj}\delta_{il}
- I^\beta_{il} \delta_{kj}) (\overline w){1\over \overline z-\overline w} -
\alpha {\delta^{il}\delta^{kj}\over (\overline z-\overline w)^2}\quad .
\eqno(4.28)
$$

For $J^\beta_{-ij}$ we are forced into a milder assumption. Indeed, since
$J^\beta_{-ij}$ is a representation of such an algebra with a central
extension and since it commutes with itself, the equation $\partial_+J^\beta_
{-ij}=0$
would be too simple to realize the whole problem we are considering. In such a
case we would be left with unequal time commutators for the last  equation
(4.10). But in any case, since $I^\beta_{ij}$ is a right-moving field operator,
the equal time requirement in the first equation (4.10) is also superfluous,
and we get an  operator product algebra of the type
$$
I^\beta_{-ij}(\overline z) J^\beta_{-kl}(w, \overline w) = (J^\beta_{-kj}\delta
_{ij} - J^\beta_{-il}\delta_{kj})(w, \overline w) {1\over \overline z-
\overline w} + 2 {\delta^{il} \delta^{kj}\over (\overline z -\overline w)^2}
\quad ,\eqno(4.29)
$$
where once again we turned to the Euclidian variables. The  consequence is the
fact that holomorphic derivatives of the current are indeed primary fields.
However the second equation in (4.10) cannot be taken at arbitrary times,
since $J^\beta_-$ depends on both $x^+$ {\it and} $x^-$. Moreover, if
$J^\beta_-$ were purely right-moving, the last equation would imply, for
unequal times, that it is a trivial operator.

Some conclusions may be drawn for $J^\beta_-$. As we stressed above, $\partial
_+J^\beta_-$ cannot be zero\newfoot{$^*$}{In the case $J^\beta_-$ is
left-moving, we expect further modifications of the commutators. See
discussion in the conclusions.}, in the full quantum theory; however, in view
of (4.29), we conclude that left(-) derivatives of this current are primary
fields\ref{22}, since
$$
I^\beta_{ij}\partial_+^nJ^\beta_{-kl} = {\partial_+^nJ^\beta_{-il}\delta_{kj}
-\partial_+^nJ^\beta_{-kl}\delta_{il}\over \bar z-\bar w}\quad . \eqno(4.30)
$$

Therefore, we expect a Kac--Moody algebra for $I^\beta(z)$, and $\partial ^n
I^\beta$ should be primary fields, depending on parameters $\overline z$.

Such an underlyning Kac--Moody structure is the most unexpected result in this
paper, since it arose out of a non-linear relation obeyed by the current, which
can be traced back to an integrability condition of the model. Moreover, the
theory has an explicit mass term -- although free massive fermionic theories as
well as some off-critical perturbations of conformally invariant theories in
two
dimensions may contain affine Lie symmetry algebras.

The current itself is now a realization of such algebra in its right-moving
sector. Indeed, we have derived the algebra (4.28) from the Poisson structure.

\vskip 2cm
\penalty-300
\centerline {\bf 5. The GKO construction}
\vskip .5cm
\nobreak
\noindent Gauged WZW theories provide a Lagrangian realization of the GKO
construction\ref{16,23}. Deleting the ``mass term" $\lambda$ in (2.19) we have
a gauged
WZW theory as  explicited in (2.7). The WZW functional is invariant under a
$G\times G$ symmetry transformation given by
$$
g(z,\overline z)\to\overline G (\overline z) g(z, \overline z)G (z) \quad
.\eqno(5.1)
$$
In general one can gauge the anomaly-free vector subgroup $H\subset G\times G$
by means of the addition of the term
$$
{1\over 4\pi}\tr \int {\rm d}^2x \left[ e^2 A_+A_- - e^2 a_+gA_-g^{-1} + ie
A_-g^{-1} \partial _+ g + ie A_+ g \partial _- g^{-1}\right] \quad .\eqno(5.2)
$$
In the QCD$_2$ case, $H$ corresponds to $G$.

Such a gauging procedure introduces constraints in the theory, as discussed by
Karabali and Schnitzer\ref{16}. In order to understand this point in more
detail, we have to consider the effect of the ghost sector. In general, ghosts
are introduced considering a gauge-fixing function ${\cal F}(A)$, and
introducing a factor
$$
\det \left( {\partial {\cal F}\over \partial A_\mu } {\partial A_\mu
\over \partial\epsilon }\right)\delta \left( {\cal F}(A)\right)\eqno(5.3)
$$
in the partition function, where $\epsilon $ is the gauge parameter. However,
 if we are to render explicit the conformal content of the theory, it is more
useful here to represent all possible chiral determinants in terms of ghost
integrals, so that the reparametrization invariance is also explicit and one
can later verify that the gauge-fixing procedure, as outlined above, and which
is more frequently used in the gauge field literature, is trivial in the sense
that one is led to a unit Faddeev--Popov determinant.

Therefore ghosts are introduced by writing determinants in terms of ghost
systems and decoupling them from the gauge fields by a chiral rotation,
a procedure which is possible in two-dimensional space-time. This is equivalent
 to writing all determinants as
$$
\det D = \E^{c_V\Gamma[U]} (\det \partial )^{c_V}\quad ,\quad
\det \overline D = \E^{c_V\Gamma[U]} (\det \overline \partial )^{c_V}\eqno(5.4)
$$
and substitute the free Dirac determinant in terms of ghosts as
$$
\eqalign{
(\det \partial)^{c_V} & =\int {\cal D}\overline b {\cal D}\overline c \,
\E^{-\tr \int {\rm d}^2 x \,\bar b  \partial \bar c}\quad ,\cr
(\det \overline \partial)^{c_V} & =\int {\cal D} b {\cal D} c \,
\E^{-\tr \int {\rm d}^2 x \, b  \overline \partial  c}\quad .\cr}\eqno(5.5)
$$
In fact the determinant of the Dirac operator does not factorize as in (5.4)
because of the regularization ambiguity. At every step, one has to assure
vector current conservation. Such determinants cancel out by changing some of
variables (as in (2.15)) but do not cancel in (2.19), from which we are led to
the contribution
$$
\int {\cal D} \overline b{\cal D}b{\cal D} c {\cal D} \overline c\, \E^{-\tr
\int {\rm d}^2x\, (b\overline \partial c + \overline b \partial \overline c)}
\quad .\eqno(5.6)
$$

Although decoupled at the Lagrangian level, such terms are essential due to
 constraints arising in the zero total conformal charge sector,
leading to BRST constraints on physical states. Such constraints are obtained
in
a system of interacting conformally invariant sectors $(g, \Sigma, b, \overline
b, c, \overline c )$ described by the partition function
$$
{\cal Z}= \int {\cal D}g {\cal D}\Sigma {\cal D} b {\cal D} \overline b
{\cal D} c {\cal D} \overline c \,\E^{-k \Gamma [g] + (c_V+k)\Gamma[\Sigma] -
\tr \int {\rm d}^2x \,(b \overline \partial c + \bar b \partial _+
\overline c)}\quad .\eqno(5.7)
$$
One can couple the system to external gauge fields $A^{^{ext}}$ and $\overline
A^{^{ext}}$ as above, or equivalently by means of the minimal
substitution\ref{16} $\partial \to D^{^{ext}} = \partial- ieA^{^{ext}}$ and
$\overline \partial \to \overline D^{^{ext}}=\overline \partial-ie\overline
A ^{^{ext}}$ with $ A^{^{ext}} ={i\over e} U_{_{ext}}^{-1}\partial U_{_{ext}}$,
 and $\overline A^{^{ext}}= {i\over e} V_{_{ext}} \overline \partial V_{_{ext}}
^{-1}$. The interaction of the fields from the WZW theory with such external
gauge fields is equivalently obtained from (2.7$b$), that is\ref{16}
$$
\eqalign{
-k \Gamma[g,A] & = -k\Gamma[U_{_{ext}} gV_{_{ext}} ] + k \Gamma[U_{_{ext}}
V_{_{ext}}] \cr
(c_V + k)\Gamma[\Sigma,A] & = (c_V+k)\Gamma[U_{_{ext}}\Sigma V_{_{ext}}] -
(c_V+k)\Gamma[U_{_{ext}}V_{_{ext}}]\cr
-\tr\int{\rm d}^2x\,[b\overline D^{^{ext}}c+\overline bD^{^{ext}}\overline c]
&=-\tr\int{\rm d}^2x\,[b V_{_{ext}}\overline \partial (V_{_{ext}}^{-1}c) +
\overline b U_{_{ext}}^{-1}\partial(U_{_{ext}}\overline c)]\quad
,\cr}\eqno(5.8)
$$
where $k$ is the central charge . In our preceding discussion $k=1$.

In the first two cases, the invariance of the Haar measure permits to change
variables as
$$
\eqalign{
\widetilde g & = U_{_{ext}}gV_{_{ext}} \quad ,\quad  {\cal D} \widetilde g =
{\cal D} g\quad ,\cr
\widetilde \Sigma & = U_{_{ext}}\Sigma V_{_{ext}} \quad ,\quad  {\cal D}\Sigma
= {\cal D} \widetilde \Sigma\quad ,\cr}\eqno(5.9)
$$
while in the last case one can do a chiral rotation, leaving back the free
ghost
system and a WZW term $c_V\Gamma[U_{_{ext}}V_{_{ext}}]$. Therefore, the $\Gamma
[U_{_{ext}}V_{_{ext}}]$ term cancels due to the balance of central charges
and the partition function does not depend on the external gauge fields. This
implies, in particular, that the functional derivative of the partition
function with respect to the external gauge fields vanish, therefore
$$
{\delta {\cal Z}(A^{^{ext}} , \overline A^{^{ext}})\over\delta A^{^{ext}}}\Big
\vert _{A^{^{ext}}, \overline A^{^{ext}}=0} = 0 = {\delta {\cal Z}(A^{^{ext}} ,
\overline A^{^{ext}})\over \delta \overline A^{^{ext}}}\Big\vert _{A^{^{ext}},
\overline A^{^{ext}}=0}  \quad ,\eqno(5.10)
$$
which are equivalent, due to the minimal coupling, to the set of constraints
$$
\left\langle kg^{-1}\partial g-(c_V+k)\Sigma^{-1}\partial\Sigma-4\pi[b,c]\right
\rangle=0=\left\langle J_g+J_\Sigma+J_{ghost}\right\rangle\quad ,\eqno(5.11a)
$$
as well as
$$
\left\langle k\overline\partial gg^{-1}-(c_V+k)\overline\partial\Sigma\Sigma^
{-1}-4\pi[\overline b, \overline c]\right\rangle=0=\left\langle\overline J_g+
\overline J_\Sigma + \overline J_{ghost}\right\rangle\quad .\eqno(5.11b)
$$

Each above current satisfies a Kac--Moody algebra with a corresponding central
charge. One can build up a BRST charge $Q$ as
$$
Q = \sum \colon c_{-n}^i \left( J_{g_n}^i+ J_{\Sigma_n}^i\right) \colon -
{1\over 2} i f^{ijk}\sum \colon c_{-n}^i b_{-m}^j c_{n+m}^k \colon\quad ,
\eqno(5.12)
$$
where the $i, j, k$ indices refer to the adjoint representation of the
symmetry group, $f^{ijk}$ the structure constants, and the mode expansion of
the fields read
$$
\eqalign{
c^i & =  \sum c_n^iz^{-n} \quad ,\cr
b^i & = \sum b_n^i z^{-n-1}\quad ,\cr
J_{g, \Sigma}^i & = \sum \left( J_{g,\Sigma}^i\right)_n z^{-n-1}\quad .\cr}
\eqno(5.13)
$$
The above charge is nilpotent: $\,Q^2={1\over 2}\{Q,Q\}=0$. This implies that
the above system is a set of first-\-class constraints (indeed a similar set of
constraints originates for $\overline Q$ resp. $\overline J,\overline b,
\overline c$).

The stress tensor can be computed in terms of such currents, and we have three
contributions, namely $T(z)=T_g(z)+T_\Sigma(z)+T^{ghost}(z)$, with the
respective central charges $c_g=c(g,k)={2kd_g\over c_g+2k}$, \quad $c_\Sigma= c
(H,-k-c_H)$, and $c_{ghost}=-2d_H$, where one supposes here that $\Sigma$ takes
values in $H \in G$. The total central charge, $c^{tot} = c_g + c_\Sigma +
c_{ghost}$ coincides with that obtained from the GKO construction for coset
space conformal theories, and the total energy tensor decomposes in terms of
the GKO stress energy tensor and a residual piece, $T'$, with zero central
charge.

Representations of $T'$ are thus trivial, and the
gauged WZW model is equivalent to
the GKO construction of $G/H$ conformal field theories. The physical subspace
is generated by a product of matter and ghost sectors, obeying the equation
$$
Q\vert {\rm phys}\rangle = 0\quad .\eqno(5.14)
$$
This solves also the problem of the sector with negative central charge, which
should not be considered separately, being coupled through the BRST condition.
Had we not such condition one would expect problems concerning negative metric
states. Therefore one cannot consider  each sector separately.

In the case of the inclusion of QCD$_2$ in such a scheme, we shall see that
there are further constraints. Although the new constraints seem to be of the
first-class type when considered alone, there is a combination that is second
class due to the cancellation of the ghost contribution. Therefore, in the case
of QCD$_2$ we have to deal with a Dirac quantization procedure of second-class
constraints!\ref{24}

However, we shall see that several interesting properties, characteristic of
the
model, as well as part of the conformal structural relations, still hold true,
and QCD$_2$ problem can be understood as an integrable perturbation of a
(very simple) GKO construction of coset space conformal field theory. We will
have a GKO construction of a very simple coset model to an off-critical
perturbation of a WZW theory by means of second-class constraints.

\vskip 2cm
\penalty-300
\centerline {\bf 6. Coupling to external gauge fields and constraints}
\vskip .5cm
\nobreak
\noindent Consider the Minkowskian effective action
$$
S_{eff} = \Gamma[\widetilde g] - (c_V+1)\Gamma[\Sigma] + \Gamma[\beta] -
{1\over 2}\lambda ^2 \int {\rm d}^2 x\, [\partial_+^{-1}(\beta^{-1}
\partial _+\beta)]^2 + S_{{\rm ghosts}}\quad .\eqno(6.1)
$$
Let us start with by first coupling the fields  $(\widetilde g, \Sigma, ghosts)
$ to external gauge fields
$$
A^{^{ext}}_-={i\over e}V_{_{ext}}\partial_-V_{_{ext}}^{-1}\quad ,\quad
A^{^{ext}}_+={i\over e}U_{_{ext}}^{-1}\partial_+U_{_{ext}}\quad. \eqno(6.1a)
$$
We find
$$
\eqalign{
S_{eff}(A)  = & \Gamma[U_{_{ext}}\widetilde gV_{_{ext}}] - (c_V+1)\Gamma[
U_{_{ext}}\Sigma V_{_{ext}}] + \Gamma[U_{_{ext}}\, {\rm ghosts} \,V_{_{ext}}]
+ \cr
& + [1-(c_V+1) + c_V]\Gamma [U_{_{ext}}V_{_{ext}}]\quad .\cr}\eqno(6.2)
$$
Invariance of the Haar measure, and vanishing of the total central charge (i.e.
vanishing coefficient of the last term above) tell us that the action does not
depend on the external gauge fields. Nevertheless, the action can also be
written as
$$
\eqalign{
S_{eff}(A)= & S_{eff}(0)-{1\over4\pi}A^{^{ext}}_+\left[ie\widetilde g\partial_-
\widetilde
g^{-1} - ie(c_V+1) \Sigma \partial _-\Sigma ^{-1} + J_-({\rm ghost}) \right]\cr
&-{1\over4\pi} A^{^{ext}}_-\left[ ie \widetilde g^{-1} \partial _+ \widetilde
g - ie(c_V+1) \Sigma ^{-1} \partial _+\Sigma + J_+ ({\rm ghost}) \right] +
{\cal
O}(A^2)\quad .\cr}\eqno(6.3)
$$

Functionally differentiating the partition function once with
respect to $A^{^{ext}}_+$ and separately with respect to $A^{^{ext}}_-$, and
putting $A^{^{ext}}_\pm =0$ we find the constraints
$$
\eqalign{
i\widetilde g\partial_-\widetilde g^{-1}-i(c_V+1)\Sigma\partial_-\Sigma^{-1}
+J_-({\rm ghosts})& \sim 0\cr
i\widetilde g^{-1}\partial_+\widetilde g-i(c_V+1)\Sigma^{-1}\partial_+\Sigma
+J_+({\rm ghosts})& \sim 0\cr}
\eqno(6.4)
$$
leading to two BRS charges $Q^{(\pm)}$ as discussed by [16], which are
nilpotent. Therefore we find two first-class constraints.

The field $A^{^{ext}}_+$ can also be coupled to the field $\beta$ instead of
$\widetilde g$, since the system $(\beta, \Sigma, {\rm ghosts})$ has also
vanishing
central charge. In such a case we have to disentangle the non-local interaction
considering instead of the third and fourth terms in (6.1) the $\beta$-action
$$
S(\beta) = \Gamma[\beta]+ \int {\rm d}^2 x\, {1\over 2}(\partial _+C_-)^2 +
\int {\rm d}^2 x \,\lambda C_-\beta^{-1} \partial _+\beta\quad .
\eqno(6.5)
$$

We make the minimal  substitution $\partial _+ \to \partial _+-ieA^{^{ext}}_+$,
repeating the previous arguments for the $(\beta, \Sigma, {\rm ghosts})$
system, and we now arrive at the constraint
$$
\beta \partial _-\beta ^{-1}+4\pi\lambda\beta C_-\beta^{-1} - i(c_V+1)\Sigma
\partial _-\Sigma^{-1} + J_-({\rm ghost}) \sim 0\quad .\eqno(6.6)
$$

One could na\"{\i}vely expect that, repeating the previous arguments the
system has
a new set of first-class constraints. But if we instead consider the equivalent
system of the first set, together with the difference of the $(-)$ currents,
namely
$$
\Omega_{ij}=(\beta \partial _-\beta^{-1})_{ij} + 4\pi \lambda (\beta C_-\beta
^{-1})_{ij} - (\widetilde g \partial _-\widetilde g^{-1})_{ij}\quad ,\eqno(6.7)
$$
one readily verifies that the above constraint cannot lead to a nilpotent
BRST charge due to the absence of ghosts. Therefore, it must be treated as a
second-class constraint, defining the field $C_-$. The Poisson algebra obeyed
by $\Omega_{ij}$ is
$$
\eqalign{
\left\{ \Omega_{ij}(t,x), \Omega_{kl}(t,y)\right\}&=(\widetilde \Omega_{il}
\delta_{kj} - \widetilde \Omega_{kj}\delta _{il})(t,x) \delta(x-y) + 2 \delta_
{il}\delta_{kj} \delta'(x-y)\cr
{\widetilde \Omega} & = \widetilde g \partial _- \widetilde g^{-1}+
\beta\partial_-\beta^{-1}+4\pi\lambda\beta C_-\beta^{-1}\quad .\cr}\eqno(6.8)
$$
(Notice the change of sign in $\widetilde \Omega$). Using the above, we can
thus define the undetermined velocities, and no further constraint is
generated.

The fact that the theory possesses second-class constraints is very
annoying, since these cannot be realized by the usual cohomology construction.
Therefore, instead of building a convenient Hilbert space, one has to modify
the dynamics, since the usual relation between Poisson brackets and commutators
is replaced by the relation between Dirac brackets and commutators.

Nevertheless, as we will see, several nice structures unravelled so far remain
untouched after such a harsh mutilation.  Indeed, we shall see that there
is a rather deep separation between the ``right" currents, obeying equations
analogous to those written so far, and the ``left" currents, which will obey a
modified dynamics, due to the second-class constraints.

As a consequence of the definition of the canonical momenta, eq. (4.14$c$), the
constraints have a phase-space formulation as
$$
\Omega_{ij}=4\pi (\beta \widetilde {\hat \Pi}^\beta)_{ij} + \partial _1\beta
\beta^{-1} - 4\pi (\widetilde g \widetilde {\hat \Pi}^{\tilde g})_{ij}-\partial
_1\widetilde g \widetilde g^{-1}\quad ,\eqno(6.9)
$$
which has been used to compute (6.8). Notice that the structure of the
right-hand side of the phase-space expression is rather simple. Indeed, the
$C_-$ field just redefines the momentum associated with $\beta$, and the above
constraints are analogous to those appearing in the description of non-Abelian
chiral bosons\ref{25}, that
is WZW theory with a constraint on a chiral current. It follows that the
Poisson algebra is very simple. Indeed, one finds\ref{25}
$$
\eqalign{
\left\{ \Omega_{ij}(x) , \Omega_{kl}(y)\right\} = & 16\pi\delta_{il}\delta_{kj}
\delta'(x^1\!-\!y^1) + 4\pi [ (4\pi \beta \widetilde {\hat \Pi}^\beta +
\beta'\beta^{^{-1}} \!+\! 4\pi \widetilde g \widetilde {\hat \Pi} ^{\tilde g}\!
 +\!\widetilde g' g^{^{-1}})_{kj}\delta _{il}-  \cr
 - & (4\pi \beta \widetilde {\hat \Pi}^\beta +
\beta'\beta^{-1} + 4\pi \widetilde g \widetilde {\hat \Pi} ^{\tilde g} +
\widetilde g' g^{-1})_{il}\delta _{kj}] \delta(x^1-y^1)\cr
= & 16\pi \delta_{il} \delta_{kj} \delta'(x^1-y^1) +  8\pi [ j_{-kj}
\delta_{il} - j_{-il} \delta_{kj} ] \delta (x^1-y^1)\quad ,\cr}\eqno(6.10)
$$
where $j_-= 4\pi \beta \widetilde {\hat \Pi}^\beta + \beta'\beta^{-1}$
satisfies the Poisson algebra
$$
\{ j_{-ij}, j_{-kl}\} = 8\pi \delta_{il} \delta_{kj} \delta'(x-y) + 4\pi
(j_{-kj} \delta_{il} - j_{-il}\delta_{kj})\delta(x-y)\quad .\eqno(6.11)
$$
The above expression defines also the $Q$-matrix
$$
Q_{ij;kl} = \left\{ \Omega_{ij}(x), \Omega_{kl}(y)\right\}\Big\vert_{equal
\quad time}\quad ,\eqno(6.12)
$$
which is not a combination of constraints, therefore no further constraint is
generated by the Dirac algorithm. The inverse of the Dirac matrix is not
difficult to comput and we have the expression\ref{25}
$$
\eqalignno{
\left( Q^{-1} \right)_{ij;kl}  = & {1\over 32\pi} \delta_{il} \delta_{kj}
\epsilon (x)+\cr
& + {1\over 64\pi} (\delta_{il} j_{jk} - \delta_{jk} j_{li})\vert x \vert+\cr
& + {1\over 128\pi} (\delta_{ia} j_{jb}-\delta_{jb} j_{ai})
(\delta_{al} j_{bk}-\delta_{bk} j_{la}) {1\over 2} x^2 \epsilon (x)+ &(6.13)\cr
& + {1\over 256\pi} (\delta_{ia} j_{jb}-\delta_{jb} j_{ai}) (\delta_{ca} j_{bd}
-\delta_{bd} j_{ac}) (\delta_{lc} j_{dk}-\delta_{dk} j_{cl}) {1\over 3} x^3
\epsilon (x) + \cdots \quad ,\cr}
$$
where $x$ is the space component of $x^\mu$.

The next step consists in replacing the Poisson brackets by Dirac
brackets. Thus we have to compute the Poisson brackets of the relevant
quantities with the constraints. We use
$$
\{A, B \}_{_{DB}} = \{ A, B \} _{_{PB}} - \{ A, \Omega_\alpha \} _{_{PB }}
Q_{\alpha\beta}^{-1} \{ \Omega_\beta, B \}_{_{PB}}\quad .\eqno(6.14)
$$

We will see that functions of
$$
J^\beta_+ = \beta^{-1}\partial _+\beta =- 4\pi \widetilde \Pi \beta +
\beta^{-1} \beta' + 4\pi \lambda C_-\quad \eqno(6.15)
$$
commute with $\Omega_\alpha$, and their Dirac brackets coincide with their
Poisson brackets.

Canonical quantization through the Dirac formulation of the $\beta$ sector is
achieved by the formulae (4.14$a,b,c$), (4.15), from which we obtain
$$
{1\over 4\pi} \beta^{-1}\partial _\pm\beta = - \widetilde {\hat \Pi}^\beta
\beta \pm {1\over 4\pi}\beta^{-1}\beta' + \lambda C_-\quad .\eqno(6.18)
$$

It is useful, in view of (6.8), to consider the combination
$$
{1\over 4\pi}\partial_-\beta \beta^{-1} = - \beta \widetilde {\hat\Pi}^\beta -
{1\over 4\pi} \beta' \beta^{-1} +\lambda\beta C_-\beta^{-1}\quad ,\eqno(6.19)
$$
or also, aiming at the expression of the constraint (6.8), which contains the
$C_-$ field, we have
$$
\beta \partial _-\beta^{-1} + 4\pi \lambda C_-\beta^{-1} = - 4\pi \beta
\widetilde {\hat \Pi}^\beta -  \beta' \beta^{-1}\quad .\eqno(6.20)
$$
Thus, in terms of phase-space variables the constraint is given by (6.9).
Using the above phase-space expressions we find
$$
\eqalign{
\left\{ J^\beta_-, \Omega \right\} &= 0\cr
\left\{ [J^\beta_-, \partial _+J^\beta_-], \Omega \right\} &= \left\{ [C_-,
\Pi_-], \Omega \right\} = 0\quad .\cr}\eqno(6.21)
$$
For $\left\{ \partial _+\partial _-J^\beta_-, \Omega \right\}$ we first have to
compute
$$
\eqalign{
\partial _+\partial _-J^\beta_- & = \partial _+^2J^\beta_- - 2 (\partial _+
J^\beta_-)'\quad ,\cr
& =4\pi\lambda^2 \beta^{-1}\partial_+ \beta - 2 (\Pi_-)'\quad .\cr}\eqno(6.22)
$$
We use the fact that $\{\Pi'_-, \Omega\}=0$ and we are left with
$$
\beta ^{-1} \partial _+\beta =  -{4\pi} \widetilde {\hat \Pi}^\beta \beta +
 \beta ^{-1}\beta' + 4\pi \lambda C_-\quad .\eqno(6.23)
$$
Using now $\{ C_-, \Omega \} =0$  we just have to consider
$$
j_{+ij} = \left( - 4\pi\widetilde {\hat \Pi}^\beta \beta +  \beta ^{-1}\beta'
\right)_{ij} \quad .\eqno(6.24)
$$
However, since $ \{ j_+ , j_-\}=0$ we have $\{ j_+, \Omega \}=0$! As a
conclusion, the Dirac algebra is the same as the Poisson algebra!! This is a
non-trivial result, because it holds in spite of the fact that due to (6.9)
the Dirac algebra obeyed by $\hat\Pi^\beta$ and $\beta$ changes drastically,
especially if we take into account the expression of the inverse Dirac matrix
(6.13), which is non-local and has an infinite number of terms!

\vskip 2cm
\penalty-3000
\centerline {\bf 7. BRST constraints in the dual case}
\vskip .5cm
\nobreak
\noindent In the duality transformation relating the $\beta$ and the $W$
fields, we also find interesting relations arising out of the constraint
structure of the theory. First let us perform a more detailed analysis of the
ghost structure. Back to the transformations defined by (3.9) we have
the factor $(\det \partial _+ \, \det \partial _-)^{c_V}$ left out, which
contributes as
$$
{\cal Z}^{{gh}'} = \int {\cal D} b'_+ {\cal D}b'_- {\cal D} c' {\cal D}
\overline c'\,\E^{- \tr \int {\rm d}^2 x\,(b'_+ \partial _-c' + b'_-\partial _+
\overline c')}\quad .\eqno(7.1)
$$

The coupling of a subset of fields to an external gauge potential written in
the
form (6.1$a$), as described in section $6$, can be made, and as usual. If such
a set has a vanishing  total
central charge, the partition function does not depend on the gauge potential,
and we are led to constraints again. With the partition function written in the
$W$ language as in (4.3), and taking into account all appropriate ghosts, we
have various self-commuting constraints. Some of them, such as
$$
\eqalign{
J_{\tilde g} - (c_V+1)J_\Sigma + J_{{\rm ghost}} & \sim 0\quad ,\cr
J_{\tilde \beta} - (c_V+1)J_\Sigma + J_{{\rm ghost}} & \sim 0\quad ,\cr}
\eqno(7.2)
$$
are the same as before, with the advantage that now $\widetilde \beta$ is a
pure WZW field, so that it can be simply identified with $\widetilde
g$, without further consequences. However, further constraints involving also
the $W$ field arise, such as
$$
J_+^{\tilde g} - (c_V+1)J_+^W + J_{+{\rm ghost}}  \sim 0\quad ,\eqno(7.3)
$$
in such a way that  we have, as a consequence, the non-trivial second-class
constraint
$$
J_+^\Sigma - J_+^W \sim 0 \quad ,\eqno(7.4)
$$
or, more explicitly,
$$
(c_V+1) \Sigma^{-1}\partial _+ \Sigma - (c_V + 1) W^{-1} \partial _+W + {1\over
\lambda} W^{-1} \partial _+ B W =0 \quad .
$$

Above, we proceeded  as in the $\beta$ formulation, but with the interaction of
the $A_-^{^{ext}}$ field with the $W$, while in the (dual) $\beta$ case we
considered $A_+^{^{ext}}$.

The phase-space expression is given in the formula
$$
\Omega ^{W,\Sigma} = - \widetilde {\hat \Pi}^WW + {1\over 4\pi} W^{-1} W' +
\widetilde {\hat \Pi} ^\Sigma\Sigma-{1\over 4\pi}\Sigma^{-1}\Sigma'\sim 0\quad
,\eqno(7.5)
$$
and resembles the $\beta$-formulation (see (6.9)). However, as intriguing as it
might appear, if we now substitute the $B$ field from the constraint (7.4) back
into the action we find a non-local term. This means that while in the
$\beta$-formulation, which  is non-local at the beginning, we end up with a
localaction after substituting back the constraint,  in the $W$-formulation,
which is local at the beginning, we end up with a non-local action; another
feature of duality in both formulations.

Keeping the Dirac algebra in mind, we substitute back the configuration-space
constraints into the action, maintaining  the phase-space structure. In such a
case, using (6.7), and (2.6), we redefine $\beta g\equiv P\, ,\,\beta=Pg^{-1}$,
 and find the effective action
$$
\eqalign{
S = & \Gamma[P] - {1\over 2\pi} g^{-1}\partial_+ g g^{-1}\partial_- g  -
{1\over 4\pi} P^{-1}\partial_+ P P^{-1}\partial_- P  + {1\over 4\pi} P^{-1}
\partial_+ P g^{-1}\partial_- g + \cr
+ & {1\over 4\pi} P^{-1}\partial_- P g^{-1}
\partial_+ g + {1\over 4\pi} \partial_- g g^{-1} \partial_+ P P^{-1} -
{1\over 2\pi} \partial_- g g^{-1} P g^{-1} \partial_+ g g ^{-1} P + \cr
+ & {1\over 2(4\pi\lambda)^2}\left[\partial _+\left\{ gP^{-1} g \partial _-
(g^{-1} P g^{-1})\right\}\right]^2 \quad .\cr}\eqno(7.6)
$$

The equation of motion/conservation law (3.6) still holds, as previously
proved. From action (7.6) we can find the equations of motion. Notice that the
final action is a WZW theory off the critical point, a principal
$\sigma$-model, and current--current-type interactions between them.

For the dual formulation a further interesting structure arises. The constraint
is now
$$
\partial _+B = -\lambda (c_V + 1) W \Sigma ^{-1} \partial _+(\Sigma W^{-1})
\quad .\eqno(7.7)
$$
Similarly to the above, we use (7.7) and (2.6) to introduce $S = W\Sigma$,
replacing the $W$-field. Interesting enough, it is now the dual formulation
that is non-local due to the presence of the $B$-field. We arrive again at the
WZW  theory for $S$, a principal $\sigma$-model term for $\Sigma$,
current--current-type interactions, and principal $\sigma$-model terms for $S$.
The latter are such that the (wrong) sign of the principal $\sigma$ term in
$\Gamma[S]$ changes, and we arrive at the WZW model with a relative minus
sign, or $\Gamma[S^{-1}]$!

However, the standard procedure to deal with the constraints is to substitute
the phase-space expressions in the Hamiltonian. But in such a case, the
constraint (6.9) does not depend upon $C_-$, and leads just to a connection
between the right-moving current of the $g$ sector, the left-moving current
being untouched by such a relation! Therefore, still in the present case, where
we witnessed the appearance of second-class constraints, their main role was to
assure the positive metric requirement, as we have seen by means of the change
of sign of the WZW action in the dual formulation.

\vskip 2cm
\penalty-4000
\centerline{\bf 8. Spectrum}
\vskip .5cm
\nobreak
\noindent
Having recognized the role played by the $\beta$-action, we pass to discuss the
spectrum of the theory. The first and huge step towards understanding the
model was taken by 't Hooft, who used the Bethe--Salpeter equation in the
large-$N$ limit to prove that the bound states form a Regge trajectory. By
adopting the light cone gauge $(A_-=0)$ and formulating the Feynman rules in
terms of light cone coordinates, the non-linear integral equation
$$
\Sigma (p) = -{ie^2\over 2\pi^2} \int {{\rm d}k_-\over k_-^2}\int {\rm d}k_+
{k_-+p_-\over M^2+(k+p)^2+(k_-+p_-) \Sigma(k+p) - i\epsilon} \quad ,\eqno(8.1)
$$
was obtained for the fermion self-energy $\Sigma(p)$, and the
$i\epsilon$-description was used in order to perform the ${\rm d}k_+{\rm d}k_-$
integrals. The infrared problems are very serious (as we readily see from the
above equations). Using an infrared cut-off $\lambda$ in order to restrict the
$k_-$ integration to $k_- \ge \lambda$, one can perform the integral above.
Since $k_-$ scales as a boost, the procedure turns out to be Lorentz-invariant.
The quark poles are pushed to infinity as $\lambda \to \infty$ displaying in a
clear fashion the confining properties of the theory. The solution is
$\Sigma(p)= {e^2\over \pi^2} {1\over p_-}$, and this leads\ref{1} to the
above-mentioned Regge behaviour.

However, the procedure has been subjected to some criticism. In particular,
 Wu\ref{8} pointed out that the principal value prescription is ambiguous
due to its non-commutative nature. Moreover, the above solution for the
self-energy function implies a tachyon for small bare electron mass, as
explicited in
$$
S_F (p)= {\not \! p + M + {e^2\over 2\pi} {\gamma_-\over p_-}\over p^2 - M^2 +
{e^2 \over \pi}}\quad .\eqno(8.2)
$$
In particular, the massless theory has such a tachyon pole!

Wu performed a Wick rotation, working in the Euclidean space, and after
rotating back to the Minkowki space found a different result for the fermion
self-energy, namely
$$
\Sigma(p) = {1\over p_-}\left\{ M^2 - p^2 - \sqrt{(M^2 -p^2)^2 + {4e^2\over
\pi}
p^2}\right\}^{1/2}\quad .\eqno(8.3)
$$

The anomalous branch cut reflects the fact that all rainbow graphs have been
tested for the Schwinger model. However, the complex light cone gauge involves
a
non-unitarity transformation and the relation between the results remained
unclear. By all means, there are works indicating that in the axial gauge
$n^\mu
 A_\mu =0 \, ,\, n^\mu n_\mu =-1$, it is inconsistent to use principal value
prescription\ref{9}.

Some authors have even speculated that QCD$_2$ may exist in two distinct
phases\ref{26}. In the large-$N$ regime (weak coupling) the gluons remain
massless,
since fermion loops do not contribute. Such is the 't Hooft phase. There would
exist also a Higgs phase, as in $U(1)$ gauge interaction (Schwinger model)
where the gauge field acquires a mass via the well-known Higgs mechanism. In
this case the $SU(N)$ symmetry would be broken to the maximal Abelian subgroup
of $SU(N)$.

Here we do not intend to provide a definite answer to such a complex question,
but some directions may be outlined from the computation we performed. Indeed,
we have an appropriate formulation to deal separately with both regimes: the
weak-coupling regime described by the $\beta$-action may be discussed
perturbatively. We will see that in the large-$N$ limit the relevant mass
parameter  is the one defined by 't Hooft, and we arrive at a possibility of
computing the exact mass spectrum, once the complicated constraint structure
is disentangled.

In order to understand the question concerning the spectrum, we first have to
know which is the mass of the simplest excitation, or the mass parameter
characterizing the theory. We thus consider the action
$$
S[\beta] = \Gamma[\beta] + {1\over 2} \lambda^2 \int {\rm d}^2 x \left[
\partial _+^{-1} (\beta^{-1} \partial _+\beta)\right]^2\quad ,\eqno(8.4)
$$
and write a background-quantum splitting for the $\beta$-field as
$$
\beta = \beta_0 \E^{i\xi}\quad ,\eqno(8.5)
$$
after which we have the background-quantum splitting of the action up to second
order in the quantum field $\xi$. However, we have to be careful since, in the
large-$N$ limit, the second term is the zeroth-order Lagrangian, from which we
suppose that the $\xi$ field acquires a mass $\mu^2$ to be computed. The WZW
term splits as
$$
\Gamma[\beta] = \Gamma[\beta_0] + {1\over 2} \int {\rm d}^2 x \,\,\beta_0^{-1}
\partial _\mu \beta_0 \, \xi \flex \partial _\nu \xi \,(g^{\mu\nu} + \epsilon
^{\mu\nu})\quad .\eqno(8.6)
$$

Using the fact that $\Gamma[\beta]$ is at the critical point, it is not
difficult to compute the $\beta_0^{-1}\partial _\mu\beta_0$ two-point function
at the one-loop order. We have the zeroth-order contribution from the second
term in (8.4), and the one-loop contribution, which leads to the result:
$$
\beta^{-1}\partial_+\beta {\lambda^2\over p_+^2} \beta^{-1}\partial_+\beta - N
{p_\mu p_\nu\over p^2}(g^{\mu\rho} + \epsilon^{\mu\rho})(g^{\nu\sigma} +
\epsilon^{\nu\sigma}) F(p)\beta^{-1}\partial_\rho\beta \beta^{-1} \partial_
\sigma \beta \quad ,\eqno(8.7)
$$
where
$$
F(p) = {1\over 2\pi} \sqrt{{p^2-4\mu^2\over p^2}}\ln {\sqrt {-p^2+ 4\mu^2} +
\sqrt{-p^2}\over \sqrt {-p^2+ 4\mu^2} - \sqrt {-p^2}} - {1\over \pi}\quad .
\eqno(8.8)
$$
For $p^2=\mu^2$, we find
$$
\beta_0^{-1}\partial_+\beta_0\, \beta_0^{-1}\partial_+\beta_0 {1\over p_+^2}
\left( \lambda ^2 - 4 N \mu^2 F(\mu^2)\right)\quad .\eqno(8.9)
$$

The zero of the two-point function contribution to the action is at
$$
\mu^2 = f e^2 N = f \left( e^{'t\,Hooft}\right)^2\quad ,\eqno(8.10)
$$
where $f$ is a numerical constant, in accordance with 't Hooft's results.

The fact that the second term in (8.4) has an extra factor of $N$ arises from
the fact that the fermion loops are suppressed by a factor $1/N$. Since the
fermion loops contribute with a WZW functional, while the $\lambda$ term stems
from the gauge field self-interaction (see eqs. (2.17-19)) the factors of $N$
are correct.
Moreover, it is exactly the given assignment that is compatible with the planar
expansion. Finally, we have to quote the fact that 't Hooft's analysis for the
bound state $\overline \psi \gamma_+\psi$ leads to a Bethe--Salpeter equation
compatible with the previous results, the methods following closely 't Hooft's
analysis.

A more detailed information about the spectrum of the theory can be obtained
from the Hamiltonian formulation. From the action
$$
S= \Gamma[\beta] + \int {\rm d}^2 x {1\over 2} (\partial_+C_- )^2 + \int
{\rm d}^2 x \lambda C_-\beta^{-1}\partial _+\beta\quad ,\eqno(8.11)
$$
we obtain the canonical momenta
$$
\eqalign{
\widetilde {\hat \Pi}^\beta & = {1\over 4\pi} \partial _0\beta^{-1} + \lambda
C_-\beta^{-1}\quad ,\cr
\Pi_-& = \partial _+C_-\quad ,\cr}\eqno(8.12)
$$
and the Hamiltonian density
$$
H = {1\over 2}\Pi_- (\Pi_--2C') - 2\pi (\widetilde {\hat \Pi}^\beta\beta)^2 -
{1\over 8\pi} (\beta^{-1}\beta')^2 + 4\pi \lambda \widetilde {\hat \Pi}^\beta
\beta C_- - 2\pi \lambda^2 C_-^2 - \lambda \beta^{-1}\beta'C_-\, .\eqno(8.13)
$$

The important currents are
$$
\eqalign{
J^\beta_+&=\beta^{-1}\partial_+\beta =-4\pi\widetilde {\hat \Pi}^\beta \beta +
\beta^ {-1} \beta' + 4\pi \lambda C_-\quad ,\cr
j^\beta_-&=\beta \partial _-\beta^{-1} + 4\pi \lambda \beta C_-\beta^{-1} =
4\pi
\beta \widetilde {\hat \Pi}^\beta + \beta'\beta^{-1}\quad ,\cr}\eqno(8.14)
$$
in terms of which the Hamiltonian density reads (notice that $j^\beta_-$ is
not related to $J_-^\beta$, eqs. (3.3) and (4.14$a$))
$$
H_\beta = -{1\over 16\pi} \left[ \left( J_+^{\beta}\right)^2 + \left( j_-^
{\beta}\right)^2 \right] - {1\over 2} \lambda J^\beta_+ C_- +
\pi \lambda^2 C_-^2 + {1\over 2} \Pi\left(\Pi_- - 2C'\right)\quad .\eqno(8.15)
$$

{}From the previously discussed constraint structure (6.9), the current $j_-$
is
related to the free right-moving fermion current $j_-^g=g\partial_- g^{-1}$,
and
we will drop it in the discussion of the spectrum for $\beta$. Moreover, from
the Sugawara construction of the Virasoro algebra, in terms of the Kac--Moody
generators, we know that the Sugawara piece $H_{_+}=\!{-1\over
16\pi}(J_+^\beta)^2$ acquires a factor $ (c_V + 1)^{-1}\, {\mathop{=}\limits^
{SU(N)}} \,(N+1)^{-1}$ in the quantum theory. The $C_-^2$ terms are not known,
since the $C_-$ equation of motion is not easily solvable. Nevertheless, in
terms of $C_-$ and its conjugate momentum the Hamiltonian is quadratic. If we
take it for granted that the zero-mode term is just the squared momentum, and
moreover neglecting the $C_-J_+$ interaction, the Hamiltonian eigenstates have
masses obeying the Regge behaviour
$$
m^2 \sim n \mu^2 \quad .\eqno(8.16)
$$

Corrections to this equation can be obtained using a large-$N$ expansion for
the field $C_-$, a procedure which is at least possible upon considering the
large-$N$ limit of (8.11).

\vskip 2cm
\penalty-3000
\centerline {\bf 9. Conclusions}
\vskip .5cm
\nobreak We have reached several aims in the present work. The first concerns
the issue of obtaining a bosonized version of QCD$_2$. In fact, this problem
has been solved long ago\ref{2,3,5}. Here we use those well-known methods in
order to rephrase this problem in terms of perturbations of a set of WZW
models.
Therefore features concerning integrability of the original theory are
rendered  much clearer, and an approach based on higher symmetries might be
envisaged\ref{27,28,29}.

However QCD$_2$ is a very complex theory. From the different results obtained
by 't Hooft on the one hand,  and by  Wu on the other hand, several authors
were led to support the idea that QCD$_2$ presents two phases, an unbroken weak
coupling limit, as described by 't Hooft, with a mesonic spectrum described by
a Regge trajectory, and a Higgs phase, corresponding to the break of $SU(N)$
symmetry to the maximal Abelian subgroup of $SU(N)$. There is no sign of such
spontaneous breakdown for vector-like theories, but this may happen in chiral
gauge theories as a consequence of the vacuum polarization. In order to be
able to deal with such a problem, dual formulations valid for different
regimes must be available. In this direction we found the alternatives
presented by the $\beta$ and $W$ fields, the first being an alternative for the
 week coupling limit, where we gave arguments to support 't Hooft's proof of
the Regge behaviour. However the $W$-field formulation is more involved, and we
could not draw any result based on firm ground. In any case, the integrability
of both formulations seems to be assured by the existence of higher
conservation laws, which are in fact very similar in both cases. Nevertheless,
a proof of the quantum integrability has only been possible in the $\beta$
formulation. Whether this is just a missing technical detail or a
sign of some new physics cannot be decided but by speculation.

The integrability of the theory is one of the strongest points in this work.
Several signs have been pointing, in the literature, to the possible
integrability of non-abelian gauge theories in two dimensions. Gorsky and
Nekrasov\ref{30} have studied the large-$N$ Calogero-type Hamiltonian systems
and found interesting relations with two-dimensional Yang-Mills theory. More
recently Fadeev and Korchemski\ref{31} found that the Lipatov model\ref{32} is
described by the spin zero limit of a spin system, which in turn is
integrable. Our integrability conditions eqs. (3.4-7) and (3.12) are at the
core of the integrability of the model, proving it. It would be interesting to
translate such conditions to Colagero-type Hamiltonian systems, as well as to
the Lipatov model, or still to Verlindes' high--energy description of strong
interactions\ref{33}.

We have to point out that gauged WZW models contain a rather non-trivial set of
constraints. As pointed out by [16], although there are non-interacting subsets
of fields at the Lagrangian level, the BRST constraints couple them. Such a
coupling is essential for the maintenance of positivity of Hilbert space, due
to some wrong sign of a part of the WZW actions. In the present case some
combinations of the constraints are second class, and the Dirac prodedure has
to be used in full detail. However, as it turns out, there is a decoupling
between the non-trivial sector described by the perturbed (off-critical) WZW
theory, and the constrained sector and the integrability condition turns out
to fulfil the same algebra for the Dirac as well as for the Poisson algebra.
It turns out that such is the Kac--Moody algebra, and one component of the
current is a realization of the Kac--Moody algebra.

Using such spliting between the off-critical $J_+$ current and the constraint
$j_-$ current, we can write the Hamiltonian in a convenient way, and relate
the squared momentum eigenstates to the Sugawara Hamiltonian eigenstates,
supporting the Regge behaviour obtained  by 't Hooft in the large--$N$ limit.
The same method does not seem to work in the $W$ formulation.

There is also a solution to our problem, which is compatible with the classical
structure of the current algebra, mentioned after eq. (4.28), namely
$$
\partial_+J_-=0\quad , \eqno(9.1)
$$
for both $J^\beta$ and $J^W$. If this is the case, we have to modify the
algebraic structure and impose (9.1) as a constraint, which in terms of the
canonical fields reads
$$
\Omega _1 = - {1\over \lambda (c_V+1)}B \sim 0  \eqno(9.2)
$$
in the $W$ case (for the $\beta$ case one must change $B\rightarrow \Pi_-$).

The time independence of such a constraint leads to a secondary constraint
$$
\Omega_2 =j_- - (c_V+1) \lambda \widetilde \Pi_B - {1\over 2\pi \lambda } B'
\sim \partial _+\partial _- J_-^W \sim 0 \quad .\eqno(9.3)
$$
There are no further constraints, and the Dirac $Q$-matrix is
$$
Q_{ij;kl} = \{ \Omega_{ij} ,\Omega_{kl} \}^{-1} =  \pmatrix{ (j_{-il}
\delta_{kj} - j_{-kj} \delta_{il})\delta(x^1-y^1) & \delta_{il} \delta_{kj}
\delta(x^1-y^1)\cr -\delta_{il} \delta_{kj} \delta(x^1-y^1) & 0\cr}
\quad .\eqno(9.4)
$$

In particular, for the conservation relations involving $J_-$, it leads to a
Kac--Moody algebra. Moreover, using the constraints we identify $I_-$ with the
current itself! Such a semi--classical reasoning misses the central term. The
presence of a Kac--Moody algebra in the QCD$_2$ would be sufficiently
astonishing, and we are not able, at the moment, to foresee either its
consequences, or even whether such a possibility can indeed be realized in the
present model, or speculate whether it may be so in some conformally invariant
phase. We intend to delve deeper into the algebraic structure of the model in a
subsequent publication.

\vskip 2cm
\penalty-3000
\centerline {\bf Appendix}
\vskip .5cm
\nobreak
\noindent
In Minkowski space,
$$
x^\mu = (x^0, x^1)\quad ,\quad  \partial _\pm = \partial _0 \pm \partial _1
\quad ,\quad x^\pm = x^0 \pm x^1\quad .\eqno(A.1)
$$
In Euclidean space
$$
x_\mu = (x_1, x_2)\, ,\,  \overline \partial = \partial _1 - i \partial_2
\equiv \partial _-^E \, ,\, \partial = \partial _1 + i \partial_2 \equiv
\partial _+^E \, ,\, z = x_1- ix_2 \, ,\, \overline z = x_1 + ix_2 \, .
\eqno(A.2)
$$
In order to translate from one space to the other, we have $x_2=ix_0$, implying
(notice the important $(-)$ sign!)
$$
\overline \partial \longleftrightarrow - \partial _-\quad ,\quad \partial
\longleftrightarrow \partial _+\quad .\eqno(A.3)
$$
Notice also that
$$
{\partial \over \partial \overline z} = {1\over 2}\overline \partial
\quad ,\quad  {\partial \over \partial z} = {1\over 2}\partial\quad .\eqno(A.4)
$$
With these conventions,
$$
F_{\mu\nu}F_{\mu\nu}  = 2F_{12}^2 = -{1\over 2}F_{z\bar z}^2\quad {\rm and}
\quad
F_{z \bar z}  = - i F_{12} + i F_{21} = - 2i F_{12}\quad . \eqno(A.5)
$$

Path integrals are always performed in Euclidean space, while in the canonical
quantization we use the Minkowski version.
\vskip 1cm
\noindent Acknowledgements: The authors would like to thank Prof. Luiz
Alvarez-Gaum\'e for a long series of discussions and several ideas that
contributed in a decisive way to the present work. This research was partially
supported by CAPES (E.A.), Brazil, under contract No. 1526/93-4, and by CNPq
(M.C.B.A.), Brazil, under contract No. 204220/77-7.  At a very early stage
a support from FAPESP, Brazil, under contract number 93/0765-1 is also
acknowledged (E.A.).
\vskip 2cm
\penalty-3500

\centerline {\bf References}
\vskip .5cm
\nobreak
\refer [[1]/G. 't Hooft, Nucl. Phys. {\bf B75} (1974) 461]

\refer[[2]/A.M. Polyakov and P.B. Wiegman, Phys. Lett. {\bf 131B} (1983) 121;
{\bf 141B} (1984) 223]

\refer[[3]/E. Witten, Commun. Math. Phys. {\bf 92} (1984) 455]

\refer[[4]/J. Wess and B. Zumino. Phys. Lett. {\bf 37B} (1971) 95;]

\refer[[5]/P. di Vecchia, B. Durhuus and J.L. Petersen, Phys. Lett. {\bf B144}
(1984) 245]

\refer[/D. Gonzales and A.N. Redlich, Phys. Lett. {\bf B147} (1984) 150; Nucl.
Phys. {\bf B256} (1985) 621]

\refer[/E. Abdalla and M.C.B. Abdalla, Nucl. Phys. {\bf B255} (1985) 392]

\refer[/J.L. Petersen, Acta Phys. Polon. {\bf B16} (1985) 271;]

\refer[[6]/V. Baluni, Phys. Lett. {\bf 90B} (1980) 407]

\refer[/P.J. Steinhardt, Nucl. Phys. {\bf B176} (1980) 100]

\refer[/D. Gepner, Nucl. Phys. {\bf B252} (1985) 481]

\refer[[7]/E. Abdalla, M.C.B. Abdalla and K. Rothe, {\it Non-perturbative
Methods in Two-\-dimen\-sional Quantum Field Theory}, World Scientific, 1991]

\refer[[8]/T.T. Wu, Phys. Rev. Lett. {\bf 71B} (1977) 142]

\refer[[9]/Y. Frishman, C.T. Sachrajda, H. Abarbanel and R. Blankenbecler,
Phys. Rev. {\bf D15} (1977) 2275]

\refer[[10]/D. Gross, Nucl. Phys. {\bf B400} (1993) 161]

\refer[/D. Gross and W. Taylor, Nucl. Phys. {\bf B400} (1993) 181; {\bf B403}
(1993) 395]

\refer[[11]/A. Dhar, G. Mandal and S.R. Wadia, Phys. Lett. {\bf B329} (1994)
15]

\refer[[12]/A.M. Polyakov, {\it Gauge Fields and Strings}, Contemporary
Concepts in Physics, vol 3, Harwood Scientific Publishing, 1987]

\refer[/A.M. Polyakov, Mod. Phys. Lett. {\bf A6} (1991) 635]

\refer[[13]/C.P. Burgess and F. Quevedo, Phys. Lett. {\bf B329} (1994) 457]

\refer[[14]/L. Alvarez-Gaum\'e, Trieste Summer School, Italy, 1993,
CERN-TH. 7036-93]

\refer[/E. Alvarez, L. Alvarez-Gaume and Y. Lozano, CERN-TH. 7204/94,
hepth/9403155 ]

\refer[[15]/K. Kikkawa and M. Yamasaki, Phys. Lett. {\bf B149} (1984) 357]

\refer[/T.H. Buscher, Phys. Lett. {\bf B194} (1987) 51; {\bf B201} (1988) 466]

\refer[/E. Alvarez and M.A.R. Osorio, Phys. Rev. {\bf D40} (1989) 1150]

\refer[/E. Kiritsis, Nucl. Phys. {\bf B405} (1993) 109]

\refer[[16]/D. Karabali and H.J. Schnitzer, Nucl. Phys. {\bf B329} (1990) 649]

\refer[[17]/L.D. Faddeev, {\it Lectures on quantum inverse scattering method}
in Nankai Lectures on Mathematical Physics and Integrable Systems, Ed. X.C.
Song, Singapore, World Scientific, 1990]

\refer[[18]/K. Pohlmeyer, Commun. Math. Phys. {\bf 46} (1976) 207]

\refer[/M. L\"uscher and K. Pohlmeyer, Nucl. Phys. {\bf B137} (1978) 46]

\refer[/H. Eichenherr and M. Forger, Nucl. Phys. {\bf 164} (1980) 528;
{\bf B282} (1987) 745]

\refer[[19]/E. Abdalla and K. Rothe, Phys. Rev. {\bf D 361} (1987) 3190]

\refer[[20]/E. Abdalla, M. Forger and M. Gomes, Nucl. Phys. {\bf B210} (1982)
181]

\refer[[21]/V.G. Knizhnik and A.B. Zamolodchikov, Nucl. Phys. {\bf B247}
(1984) 83]

\refer[[22]/A.A. Belavin, A.M. Polyakov and A.B. Zamolodchikov, Nucl. Phys.
{\bf B241} (1994) 333]

\refer[[23]/P. Goddard, A. Kent and D. Olive, Phys. Lett. {\bf B152} (1985)
88; Commun. Math. Phys. {\bf 103} (1986) 105.]

\refer[[24]/P.A.M. Dirac, {\it Lectures on Quantum Mechanics}, Yeshiva Univ.
Press, N.Y., 1964; Can. J. Math. {\bf 2} (1950) 129]

\refer[[25]/E. Abdalla and M.C.B. Abdalla, Phys. Rev. {\bf D40} (1989) 491]

\refer[[26]/A. Patrasciou, Phys. Rev. {\bf D 15} (1977) 3592]

\refer[/P. Mitra and P. Roy,  Phys. Lett. {\bf 79B} (1978) 469]

\refer[[27]/G. Sotkov and M. Stanishkov, {\it Off-critical $W_\infty$ and
Virasoro algebras as dynamical symmetries of integrable models}, NATO Workshop
on Integrable Quantum Field Theories, Como, Italy, 1992]

\refer[[28]/E. Abdalla, M.C.B. Abdalla, G. Sotkov and  M. Stanishkov, Int. J.
Mod. Phys. A, to appear]

\refer [[29]/E. Abdalla, M.C.B. Abdalla, J.C. Brunelli and A. Zadra, Commun.
Math. Phys. to appear]

\refer[[30]/A. Gorsky and N. Nekrasov, Nucl. Phys. {\bf B414} (1994) 213]

\refer[[31]/L.D. Faddeev and G.P. Korchemsky, Leningrad preprint ITP-SB-94-14,
hepth/9404173]

\refer[[32]/L.N. Lipatov, Phys. Lett. {\bf B251} (1990) 284; {\bf 309} (1993)
394; Sov. Phys. JETP {\bf 63} (1986) 904]

\refer[/Ya. Balitsky and L.N. Lipatov, Sov. J. Nucl. Phys. {\bf 28} (1978) 822]

\refer[[33]/E. Verlinde and H. Verlinde, Princeton preprint PUPT-1319, Sept.
93]
\end